\documentclass[AMA,Times1COL]{WileyNJDv5}

\articletype{Special Issue: Perspective}

\received{November 1, 2023}
\revised{\today}
\accepted{}
\journal{J Softw Evol Proc}
\volume{00}
\copyyear{\the\year}
\startpage{1}

\usepackage{array}
\usepackage{circledsteps}
\usepackage{enumitem}
\usepackage{lastpage}
\usepackage{mdframed}

\usepackage[authormarkup=none]{changes}
\definechangesauthor[color=blue]{EC}
\definechangesauthor[color=violet]{TB}
\definechangesauthor[color=blue]{FO}

\newcommand{\blue}[1]{\textcolor{blue}{#1}}

\graphicspath{{figures/}}

\newcommand{\pdoi}[1]{\href{https://doi.org/#1}{doi: #1}}

\begin{document}

\title{Looking back and forward: A retrospective and future directions on Software Engineering for systems-of-systems}

\author[1]{Everton Cavalcante}
\author[1]{Thais Batista}
\author[2]{Flavio Oquendo}

\authormark{CAVALCANTE \textsc{et al.}}
\titlemark{\MakeUppercase{Looking back and forward: A retrospective and future directions on Software Engineering for systems-of-systems}}

\address[1]{\orgname{Federal University of Rio Grande do Norte}, \orgaddress{\city{Natal}, \country{Brazil}}}

\address[2]{\orgdiv{IRISA-UMR CNRS}, \orgname{Université Bretagne Sud}, \orgaddress{\city{Vannes}, \country{France}}}

\corres{Corresponding author Everton Cavalcante, Instituto Metrópole Digital, Universidade Federal do Rio Grande do Norte, Campus Universitário - Lagoa Nova, 59078-900, Natal, Rio Grande do Norte, Brasil. \email{everton.cavalcante@ufrn.br}}


\abstract[Abstract]{Modern systems are increasingly connected and more integrated with other existing systems, giving rise to \textit{systems-of-systems} (SoS). An SoS consists of a set of independent, heterogeneous systems that interact to provide new functionalities and accomplish global missions through emergent behavior manifested at runtime. The distinctive characteristics of SoS, when contrasted to traditional systems, pose significant research challenges within Software Engineering. These challenges motivate the need for a paradigm shift and the exploration of novel approaches for designing, developing, deploying, and evolving these systems. The \textit{International Workshop on Software Engineering for Systems-of-Systems} (SESoS) series started in 2013 to fill a gap in scientific forums addressing SoS from the Software Engineering perspective, becoming the first venue for this purpose. This article presents a study aimed at outlining the evolution and future trajectory of Software Engineering for SoS based on the examination of 57 papers spanning the 11 editions of the SESoS workshop (2013-2023). The study combined scoping review and scientometric analysis methods to categorize and analyze the research contributions concerning temporal and geographic distribution, topics of interest, research methodologies employed, application domains, and research impact. Based on such a comprehensive overview, this article discusses current and future directions in Software Engineering for SoS.}


\jnlcitation{\cname{
\author{Cavalcante E},
\author{Batista T},
\author{Oquendo F}}.
\ctitle{Looking back and forward: A retrospective and future directions on Software Engineering for systems-of-systems.} \cjournal{\it J Softw Evol Proc.} \cvol{\the\year;00(00):1--\pageref{LastPage}}.}

\frenchspacing
\maketitle

\renewcommand\thefootnote{}
\footnotetext{\textbf{Abbreviations:} ACM CCS, ACM Computing Classification System; CBSoft, Brazilian Conference on Software: Theory and Practice; ECSA, European Conference on Software Architecture; ICSA, IEEE International Conference on Software Architecture; ICSE, IEEE/ACM International Conference on Software Engineering; INCOSE, International Council on Systems Engineering; MAPE-K, Monitor--Analyze--Plan--Execute over a shared Knowledge; MDE, Model-Driven Engineering; SBES, Brazilian Symposium on Software Engineering; SESoS, International Workshop on Software Engineering for Systems-of-Systems and Software Ecosystems; SLR, systematic literature review; SMC: statistical model checking; SMS: systematic mapping study/studies; SoIS, system(s) of information systems; SoS, system(s)-of-systems; SoSE, IEEE International Conference on System of Systems Engineering; VV\&T: Verification, Validation, and Testing; WDES, Workshop on Distributed Software Development, Software Ecosystems and Systems-of-Systems.}

\renewcommand\thefootnote{\fnsymbol{footnote}}
\setcounter{footnote}{1}

\section{Introduction}
\label{sec:introduction}
Current systems have become increasingly complex mainly due to the diverse hardware and software resources they require, the processes (processing, monitoring, management, communication, etc.) involved in their life cycle, and the interaction with other systems, actors, and artifacts. In this context, it has been possible to observe a growing interest in the research and development of large-scale complex systems resulting from integrating other independent systems, the so-called \textit{systems-of-systems} (SoS). This class of systems has been observed in several application domains, such as defense, environmental monitoring, energy, emergency management and coordination, transportation, healthcare, and smart cities. In these scenarios, it is common to have high-level objectives that cannot be addressed separately by the elements that make up the system, but that can eventually be satisfied as a result of the integration and interaction between them.\cite{Cadavid2020}

SoS represents a topic in evident research activity, and several studies in the literature present different definitions and characterizations for SoS.\cite{Nielsen2015} Recently, the ISO/IEC/IEEE 21840:2019 International Standard defined SoS as ``a set of systems and system elements that interact to provide a unique capability that none of the constituent systems can achieve on its own.\cite{ISO21840}'' This is one of the most notable characteristics of SoS, in that the interaction between constituent systems makes an SoS able to offer new functionalities not provided by any of those constituents if they operate in isolation. Furthermore, the community has achieved convergence in a set of characteristics inherent to this class of systems, \textit{all} of them to be satisfied to consider a system as an SoS.\cite{Maier1998} These characteristics, which distinguish an SoS from other complex and large-scale systems, are:

\begin{itemize}
\item \textit{Operational independence}: Every constituent system of an SoS operates independently of each other to fulfill its own mission;
\item \textit{Managerial independence}: Every constituent system of an SoS is managed independently and may decide to evolve in ways unforeseen when they were initially combined;
\item \textit{Geographical distribution}: Constituent systems of an SoS are physically decoupled in the sense that only information is transmitted among constituent systems;
\item \textit{Evolutionary development}: As a consequence of the independence of the constituent systems, an SoS as a whole may evolve to respond to changes in its constituent systems and operational environment and sustain its missions;
\item \textit{Emergent behavior}: The behavior of an SoS emerges from the local interaction of its constituent systems and cannot be performed by any constituent system alone. This behavior may be ephemeral because the systems composing the SoS evolve independently, thus impacting their availability.
\end{itemize}

\added[id=FO]{As an example of an SoS, consider the case of autonomous drones owned by city councils in a metropolitan area. Each drone has its own mission assigned by its owning city council. They may collectively form an SoS to fulfill a mission that would not be feasible by a drone alone. For instance, the available drones could be deployed to participate in a reconnaissance mission for flood monitoring and early warning. In this situation, those drones must get together and fly to the zone under risk, typically as a flock, i.e., as a clustered group of drones moving with a standard velocity. Once they arrive at the target zone, they naturally split into smaller flocks, each aiming to monitor a smaller area inside the zone. Therefore, they must create and maintain flocks, split into smaller ones, and join to form larger new flocks. This reconnaissance SoS complies with the five distinguishing characteristics of SoS in the following:}

\begin{itemize}
\item \added[id=FO]{\textit{Operational independence:} Each drone operates independently of each other to fulfill missions assigned by its owning city council.}
\item \added[id=FO]{\textit{Managerial independence:} Each drone is independently managed by its owning city council.}
\item \added[id=FO]{\textit{Geographical distribution:} All the drones are physically dissociated.}
\item \added[id=FO]{\textit{Evolutionary development:} Due to the independence of the drones, the reconnaissance SoS is evolutionarily developed. Each drone may have evolving capabilities, new drones may be acquired, and existing drones may be withdrawn from service.}
\item \added[id=FO]{\textit{Emergent behavior:} The flocking of autonomous drones, required for fulfilling the reconnaissance mission of the SoS, is an emergent behavior formed and maintained by the participating drones, i.e., the flockmates, including splitting and joining. Note that there are no predefined relationships in flocking, and flockmates may autonomously change their position within the flock at any time.}
\end{itemize}

\added[id=FO]{The most notable of these characteristics is emergent behavior, which, in the case of the reconnaissance SoS, needs to create and maintain the drone flocking. As a flockmate, each drone uses its sensors to acquire information about its environment and neighbors (flockmates outside its local neighborhood are ignored) and acts according to its relative position regarding its neighbors. A question arising in this scenario is: how can a completely decentralized SoS, composed of autonomous drones, yield emergent behavior, even when there is no communication among them? In other words, how can each drone autonomously determine the acceleration vector for its own steering maneuver for collectively creating and maintaining flocking? Moreover, note that this aerial reconnaissance SoS operates in an open-world environment, i.e., its operational environment, a flood zone, is only partially known at design time, and there will inevitably be novel situations to deal with at runtime. The drones participating in this SoS are only partially known at design time as the available drones and their capabilities may change due to the independence of the city councils. Issues related to engineering software-intensive SoS, as exemplified, are intrinsically different from those of engineering single systems, thereby requiring a dedicated body of knowledge for SoS.}

\added[id=EC]{The mainstream body of knowledge of Software Engineering primarily focuses on single systems, not on SoS.\cite{Oquendo2016a} Many software engineers are used to designing and developing centralized or distributed software, but not decentralized, as is the case for SoS. In the software-intensive systems traditionally addressed by Software Engineering, constituent parts (components or sub-systems) are typically known and assembled to fulfill missions and offer functionalities defined at design time, operating in a well-defined, closed-world environment. SoS do not feature these characteristics since their constituents are operationally and managerially independent systems that might be only partially known at design time and dynamically discovered to contribute to delivering more complex functionalities. The main difference between an SoS and a single system is undoubtedly the nature of their constituent systems (specifically their level of independence) and the exhibition of emergent behavior.}

\added[id=FO]{Another critical complicating factor for SoS is that they are designed to produce an emergent behavior on which they rely for fulling missions, unlike traditional systems. This leads to the following oxymoron: how to design and develop an SoS to exhibit expected (by design) unexpected (emergent) behavior? By definition, an emergent behavior is unexpected, i.e., the behavior of a whole (the SoS) that cannot be predicted exclusively from its parts (the constituent systems) even they are known. This means that the behavior of an SoS is more than the sum of the behavior of its constituent systems. For the case of the drone flocking, an external observer of the reconnaissance SoS will naturally perceive the flocking behavior when the drones fly together (like in natural systems, e.g., flock of birds, herds of land animals, and schools of fishes). However, s/he will not be able to reduce that emergent behavior to the individual behaviors of the constituent systems, i.e., the individual drones. On the other hand, an external observer of the individual drones, even if s/he knows the behavior of each individual drone, will not be able to deduce the emergent behavior of flocking.}

\added[id=EC]{SoS have been studied for several years within Systems Engineering and still represent a relevant research topic, mainly considering the ever-increasing complexity and scale of contemporary systems and the impact these systems can bring to several application domains. The trend of software evolution and the unprecedented reliance of economy and society upon software-intensive systems have led many SoS to be fundamentally driven by software, i.e., being software-intensive as well.\cite{Oquendo2016a,Goncalves2014} On the other hand, the characteristics that make SoS unique are the same ones that pose significant challenges to both Systems Engineering and Software Engineering. Some researchers have acknowledged that traditional approaches are limited or cannot address the peculiarities and complexity of SoS, which are often misunderstood, underestimated, or reduced to a simplistic point of view.\cite{Azani2009,Lana2016}}

\added[id=EC]{The literature has also emphasized the need for a paradigm shift in the life cycle of software-intensive SoS, implying breakthrough changes in requirements engineering,\cite{Lewis2009,Lana2019} software architecture,\cite{Cadavid2020,Oquendo2016a} development,\cite{GracianoNeto2018} verification, validation, and testing (VV\&T),\cite{Luna2013,OliveiraNeves2018} quality assurance,\cite{Nelson2020} and deployment.\cite{Dakkak2023} The many challenges and open issues in this context motivate the existence of forums where researchers and professionals can exchange ideas and experiences, analyze research and development issues, and discuss solutions in Software Engineering for SoS.}

The \textit{IEEE International Conference on System of Systems Engineering} (SoSE)\footnote{\url{http://www.sosengineering.org/}}, held for almost two decades by the IEEE System, Man, and Cybernetics Society and the International Council on Systems Engineering (INCOSE), is an international conference whose interest lies in a wide range of topics related to theories, methods and applications in Systems Engineering. Although there certainly are scientific contributions dealing with Software Engineering for SoS, the SoSE conference does not have this background. Due to the lack of a specific forum about this field, the \textit{International Workshop on Software Engineering for Systems-of-Systems} (SESoS) emerged in 2013 with a focus on the theory and practice of how Software Engineering could be explored to construct software-intensive SoS.

SESoS is a workshop that has been held annually as part of important Software Engineering international conferences, such as the IEEE/ACM International Conference on Software Engineering (ICSE), the IEEE International Conference on Software Architecture (ICSA), and the European Conference on Software Architecture (ECSA). The workshop has focused on various topics related to the design, construction, analysis, and evolution of software-intensive SoS, in addition to being interested in reporting experiences and successful cases related to SoS and discussing research challenges and future perspectives in Software Engineering for SoS. With these characteristics, SESoS became the first forum to address SoS from a Software Engineering perspective, annually attracting researchers and professionals from different locations around the world. As an evolution of the research field and the workshop, SESoS 2022 revised and expanded its scope to address systemic thinking and complexity in SoS and software ecosystems, making it possible to reflect on how the different dimensions of these systems (technological, organizational and social) must be addressed in research and practice.\cite{Santos2023} A consequence of this change was renaming the workshop to \textit{International Workshop on Software Engineering for Systems-of-Systems and Software Ecosystems} since 2022.

As a result of the 11 editions held to date, the volume of scientific material documented in SESoS papers is significant for analyzing the nature, extent, and impact of research in Software Engineering for SoS. From this perspective, this article reports a study aimed at understanding the past, present, and future of Software Engineering for SoS through an overview of the research reported in papers published at the SESoS workshop between 2013 and 2023. The study combined scoping review\cite{Munn2018,Campbell2023} and scientometric analysis\cite{VanRaan1997,Mingers2015} methods to categorize and analyze research in Software Engineering for SoS presented by a corpus of 57 papers available at the SESoS proceedings. 

\added[id=EC]{The main contributions of this article are twofold. First, it offers a comprehensive landscape of what has been done regarding Software Engineering for SoS by analyzing the research reported in the SESoS workshop. This analysis delves into various aspects, including the temporal distribution and geographic location of research, prominent topics, research methodologies employed, application domains tackled, addressed SoS types, and research impact. Second, by identifying research gaps within this landscape, this article highlights topics requiring further investigation in Software Engineering for SoS.}

The remainder of this article is structured as follows. Section~\ref{sec:sesos} presents a brief motivation and history of the SESoS workshop. Section~\ref{sec:methodology} describes the methodology used in this study, including the research methods, questions, and adopted procedures. Section~\ref{sec:results} contains a detailed analysis of the SESoS research contributions, discussed from different facets. Section~\ref{sec:directions} highlights current and future directions in Software Engineering for SoS. Finally, Section \ref{sec:conclusion} contains some conclusions.

\section{The SESoS workshop}
\label{sec:sesos}
The SESoS workshop series was created in 2013 to provide a forum for researchers and practitioners to exchange ideas and experiences, analyze research and development issues, discuss promising solutions, and propose visions for the future in Software Engineering for SoS. At that time, SoS were becoming a hotspot from both the research and the industrial viewpoints, in particular related to crucial applications, such as safety-critical or mission-critical SoS. However, there was a lack of forums addressing SoS from the Software Engineering perspective. SESoS was the first venue specifically addressing Software Engineering for SoS.

The first edition of SESoS was organized in Montpellier, France, as a one-day, full-day workshop within ECSA, one of the most relevant international conferences about software architectures, attracting participants from different places worldwide. Since then, two categories of papers have been typically solicited, namely (i) regular research papers describing original, substantial contributions in research and practice and (ii) short/position papers and experience reports describing emerging results, lessons learned, open problems, novel ideas, and perspectives on research. The workshop program has been annually organized upon invited keynotes and technical sessions for paper presentations and discussions among participants. The topics of interest of SESoS have covered different aspects of SoS development from a Software Engineering perspective, including architectural design, development techniques, extra-functional concerns, and verification and validation. The workshop has also welcomed contributions reporting experiences, successful case studies, and papers discussing open issues, challenges, and future perspectives for future research. \tablename~\ref{tab:sesos-topics} summarizes the general topics of interest of the SESoS workshop.

\begin{center}
\begin{table*}
\caption{General topics of interest of the SESoS workshop.\label{tab:sesos-topics}}
\begin{tabular*}{\textwidth}{@{}lp{14cm}@{}}
\toprule
Category & Topics\\
\midrule
Analysis \& Architecture & Requirements Engineering, architectural description, architectural evaluation, architectural evolution\\
Model-Based Engineering & Modeling and simulation, Model-Driven Engineering, models at runtime\\
Construction \& Evolution & Techniques and technologies, service-orientation, component framework, evolutionary development\\
Experience & Industrial perspectives, successful case studies\\
General Issues & Characteristics of SoS, quality in SoS, SoS development processes, acquisition and project management, verification and validation of SoS\\
Challenges \& Directions & Future perspective, challenges, and directions of SoS research and development\\
\bottomrule
\end{tabular*}
\end{table*}
\end{center}

\begin{figure*}
\centerline{\includegraphics[width=0.85\textwidth]{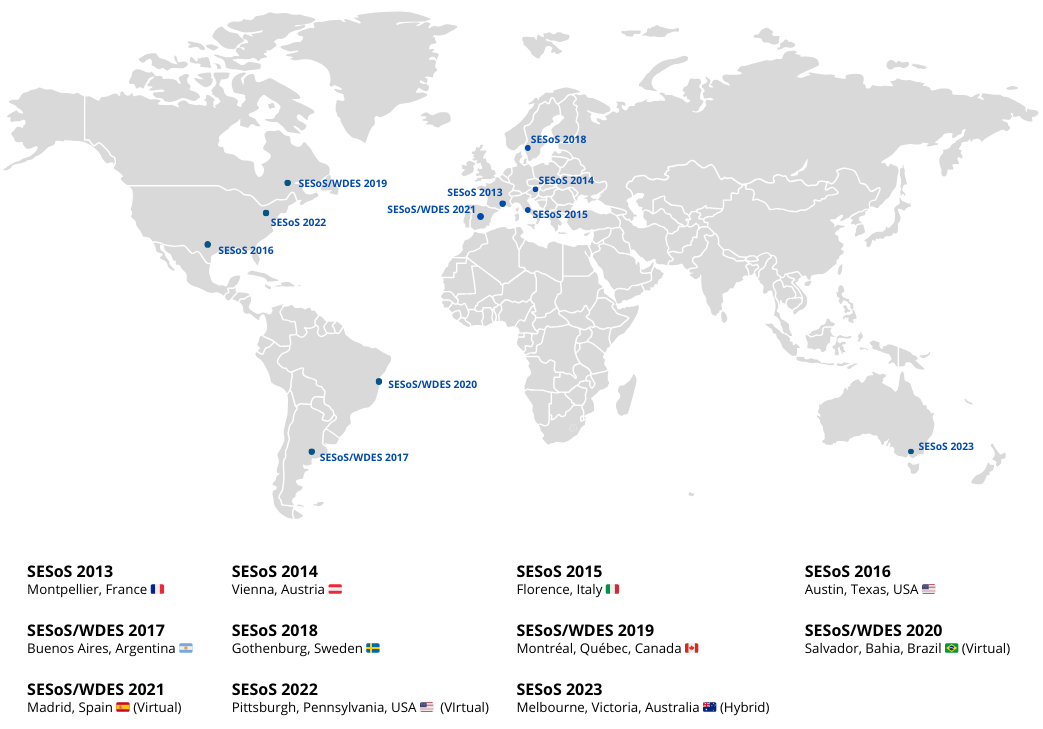}}
\caption{Locations of SESoS editions (2013-2023).\label{fig:sesos-editions}}
\end{figure*}

\figurename~\ref{fig:sesos-editions} illustrates a world map showing the location of all SESoS editions. Eleven editions of SESoS have been held so far in conjunction with ECSA, ICSA, and ICSE. The co-location of SESoS with premier venues on Software Engineering has attracted researchers and practitioners working on or interested in approaching SoS, besides contributing to a greater international visibility of the workshop. The two first SESoS editions (2013 and 2014) were organized as part of ECSA, and eight editions have been held yearly with ICSE since 2015. In 2020, the workshop was part of the ICSA program. \tablename~\ref{tab:sesos-editions} presents some historical information about the SESoS editions. The upcoming SESoS edition, the 13th of the series, is planned to happen again with ICSE in May 2024: \url{http://sesos2024.icmc.usp.br}.

\begin{center}
\begin{table*}
\caption{SESoS editions.\label{tab:sesos-editions}}
\begin{tabular*}{\textwidth}{@{}p{3cm}p{2.5cm}p{2cm}p{6.5cm}c@{}}
\toprule
Edition & Date & Venue & URL & Published papers\\
\midrule
SESoS 2013 & July 2, 2013 & ECSA 2013 & \url{http://sesos2013.icmc.usp.br} & 10\\
SESoS 2014 & August 29, 2014 & ECSA 2014 & \url{http://sesos2014.icmc.usp.br} & 11\\
SESoS 2015 & May 17, 2015 & ICSE 2015 & \url{http://sesos2015.icmc.usp.br} & 8\\
SESoS 2016 & May 16, 2016 & ICSE 2016 & \url{http://sesos2016.icmc.usp.br} & 6\\
SESoS/WDES 2017	& May 23, 2017 & ICSE 2017 & \url{http://sesos-wdes-2017.icmc.usp.br} & 15\\
SESoS 2018 & May 29, 2018 & ICSE 2018 & \url{http://sesos2018.icmc.usp.br} & 5\\
SESoS/WDES 2019 & May 28, 2019 & ICSE 2019 & \url{http://sesos-wdes-2019.icmc.usp.br} & 10\\
SESoS/WDES 2020	& November 2, 2020 & ICSA 2020	& \url{http://sesos-wdes-2020.icmc.usp.br} & 8\\
SESoS/WDES 2021	& June 3, 2021 & ICSE 2021 & \url{http://sesos-wdes-2021.icmc.usp.br} & 5\\
SESoS 2022 & May 16, 2022 & ICSE 2022 & \url{http://sesos2022.icmc.usp.br} & 6\\
SESoS 2023 & May 14, 2023 & ICSE 2023 & \url{http://sesos2023.icmc.usp.br} & 10\\
\hline
\multicolumn{4}{@{}l}{Average} & 8.55\\
\bottomrule
\end{tabular*}
\end{table*}
\end{center}
 
Four editions of SESoS (2017, 2019, 2020, and 2021) were held jointly with the Workshop on Distributed Software Development, Software Ecosystems, and Systems-of-Systems (WDES). The WDES workshop series started in 2007 in Brazil and was co-located with the leading national Software Engineering conferences, the Brazilian Symposium on Software Engineering (SBES) and later the Brazilian Conference on Software: Theory and Practice (CBSoft). WDES initially focused on distributed software development, an approach based on distributed resources and personnel to reduce cost, improve productivity and software quality, and reach new Information Technology markets. The workshop evolved over the years to include software ecosystems and SoS, which have Software Engineering-related problems and challenges amplified due to their distributed nature. A software ecosystem refers to a system formed over a common technological platform that joins a set of actors and artifacts that are responsible for software solutions based on contributions from a socio-technical network\cite{Santos2024} so that the social and technical issues related to software development made this topic as of necessary investigation in Software Engineering. WDES gained international projection by joining SESoS in ICSE 2017 and when it was held in conjunction with ECSA 2018.

The successful four editions of SESoS/WDES as a joint workshop and the synergy between SoS and software ecosystems resulted in merging SESoS and WDES as a single workshop in 2022, when SESoS was renamed to \textit{International Workshop on Software Engineering for Systems-of-Systems and Software Ecosystems}, now explicitly incorporating software ecosystems. Putting SoS and software ecosystems together enables researchers and practitioners to understand how different intertwined dimensions (technical, social, business, etc.) affect the quality and evolution of these systems.\cite{Santos2024} Therefore, SESoS took a step forward to cope with the challenges and opportunities in the design, development, deployment, and evolution of complex software-intensive systems such as SoS and software ecosystems, now considering technological, organizational, and social aspects in this context.\cite{Santos2023}

\section{Methodology}
\label{sec:methodology}
\added[id=EC]{The research methodology adopted in this work combined \textit{scoping review} and \textit{scientometric analysis}. These methods are briefly described in the following.}

\textbf{Scoping review.}
A \textit{scoping review} is a type of secondary study that seeks to synthesize evidence by identifying and evaluating the literature on a specific topic, field, concept, or issue, primarily to establish an overview or categorization of existing work in a given area of research, irrespective of type (primary studies, secondary studies, non-empirical evidence). \added[id=EC]{A critical strength of scoping reviews is that they can provide a rigorous method for mapping a field of interest regarding the volume, nature, and characteristics of research in a time relatively shorter than systematic literature reviews (SLRs) and systematic mapping studies (SMS).\cite{Arksey2005}}

Due to their aim of understanding the nature and extent of available research evidence, scoping reviews tend to be more exploratory than SLRs and SMS.\cite{Campbell2023,Grant2009} It is worth mentioning that scoping review is a term that is sometimes used in a similar way or even interchangeably with SMS (in Software Engineering as well\cite{Ralph2022,Kitchenham2023}) since both methods have much in common regarding objectives and procedures.\cite{Arksey2005,Pham2014} However, this study follows Campbell et al.\cite{Campbell2023} in considering scoping reviews and SLRs/SMS as distinct concepts.

Scoping reviews are suitable for categorizing bodies of knowledge in complex, heterogeneous areas\cite{Peters2020} like Software Engineering for SoS. This study aimed to categorize research in Software Engineering for SoS presented in SESoS papers concerning time, geographic location, topics of interest, types of research conducted, and application domains or contexts.

\begin{figure*}
\centerline{\includegraphics[width=\textwidth]{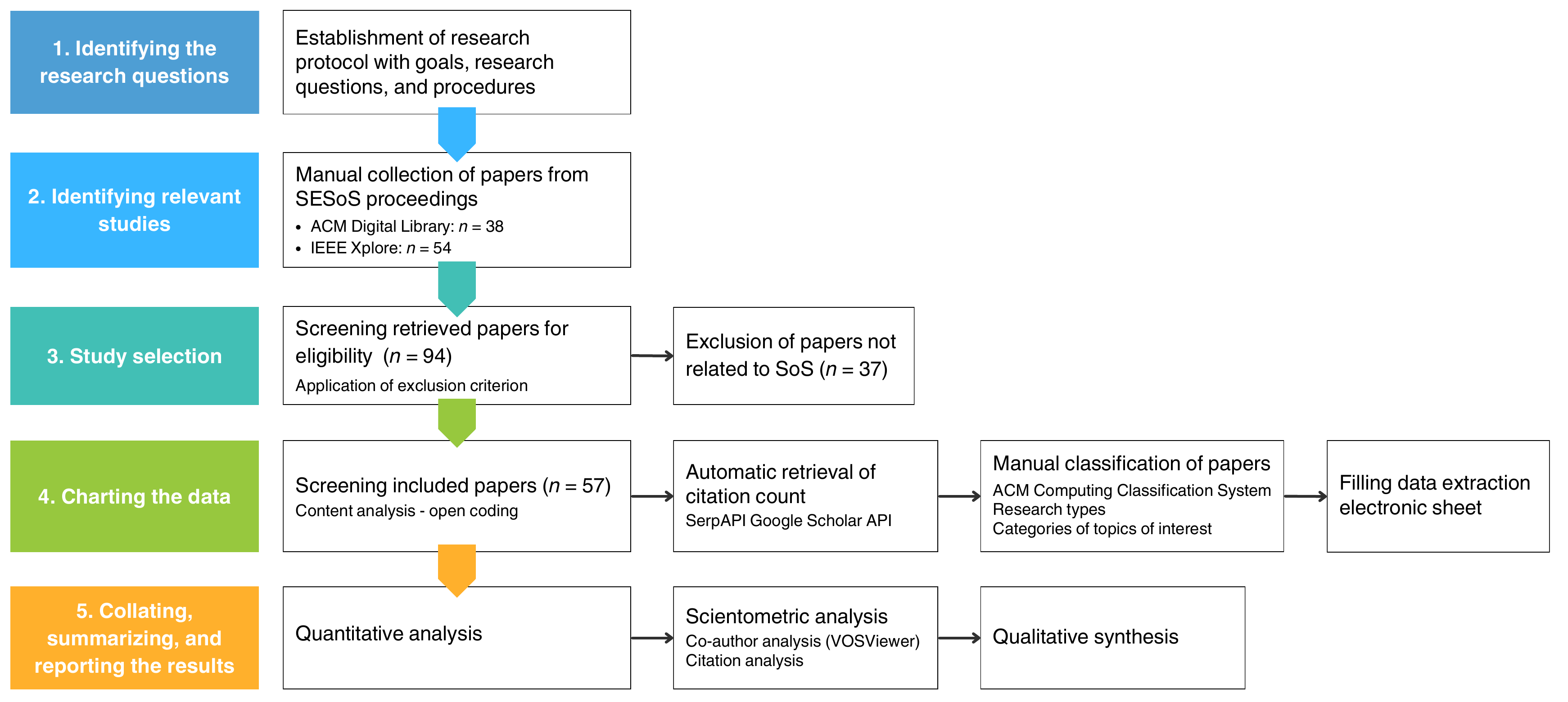}}
\caption{Process for retrieving and analyzing the SESoS papers.\label{fig:process}}
\end{figure*}

\added[id=EC]{\figurename~\ref{fig:process} depicts the process for retrieving and analyzing the SESoS papers using the scoping review method. The study followed the guidelines outlined by Arksey and O'Malley\cite{Arksey2005} and Peters et al.\cite{Peters2020} as a sequence of five stages: \Circled{1} \textit{identifying the research questions} to be answered aligned to the study goals, all documented into a protocol to support transparency and minimize bias; \Circled{2} \textit{identifying relevant studies}, which consists in comprehensively searching for studies from different sources in the literature; \Circled{3} \textit{study selection}, which stands for the application of the inclusion and exclusion criteria defined in the protocol while screening the retrieved studies to filter out those relevant to answer the research questions; \Circled{4} \textit{charting the data}, which refers to extracting and interpreting data from the selected studies; and \Circled{5} \textit{collecting, summarizing, and reporting the results} as an overview of the reviewed literature, primarily relying on qualitative synthesis.}

\textbf{Scientometric analysis.}
Scientometrics refers to measuring aspects of scientific activities, such as performance, impact, international collaboration, and others, based on the academic production of a given field, i.e., papers and citations.\cite{VanRaan1997} In recent years, scientometrics has come to play a significant role in the measurement and evaluation of research performance and mapping scientific knowledge.\cite{Mingers2015}

Scientometric analysis typically uses four techniques: (i) \textit{co-word analysis}, which comprises building networks of co-occurring terms or keywords; (ii) \textit{co-author analysis}, which focuses on collaboration networks established among authors, institutions, and countries; (iii) \textit{citation analysis}, which concern analyzing the number of citations of papers and venues as measures of impact; and (iv) \textit{document clustering}, which reveals similarities between papers. \added[id=EC]{In this study, scientometric analysis based on co-author and citation analysis techniques complemented the fifth stage of the scoping review (see \figurename~\ref{fig:process}).}

The remainder of this section details some aspects of the study methodology. Section~\ref{subsec:rqs} poses the research questions guiding this study. Section~\ref{subsec:search+selection} describes the procedures for collecting SESoS papers and filtering out the ones for further analysis. Section~\ref{subsec:dataextraction+analysis} details the data extraction and analysis procedures.

\subsection{Research questions}
\label{subsec:rqs}
This study aims to understand the past, present, and future of Software Engineering for SoS by providing an overview of the research reported in the papers that appeared in the SESoS workshop. To achieve this goal, six research questions (RQs) were formulated as follows:

\begin{enumerate}[label={RQ\arabic*:},leftmargin=*,itemsep=4pt]
    \item Where research related to Software Engineering for SoS has been conducted?\\
    \textit{Rationale:} Identifying the SESoS authors' origins based on their affiliations can reveal where research is concentrated, how the community collaborates, and possible interactions in academia and between academia and industry.
    
    \item What are the most prevalent topics addressed by the papers?\\
    \textit{Rationale:} The classification of each paper into the topics listed in the workshop's Call for Papers allows identifying the most addressed topics and understanding how the interest in them has changed over time.
    
    \item Which Software Engineering fields do the papers intersect with?\\
    \textit{Rationale:} The classification of each paper into a taxonomy of Software Engineering topics allows mapping the research on Software Engineering for SoS to a broader field.

    \item Which types of research do the studies concern?\\
    \textit{Rationale:} The classification of each paper into a research type allows for identifying the proposed approaches and assessing the degree of maturity of the contributions.

    \item What application domains or contexts are the studies inserted into?\\
    \textit{Rationale:} Identifying the application domains or contexts can reveal if the research reported in SESoS papers is specific or more generic.

    \item \added[id=EC]{Which SoS types do the studies concern?\\
    \textit{Rationale:} Identifying the SoS types the studies concern is crucial to understanding the extent of the research reported in SESoS papers.}

    \item How is the impact of SESoS papers?\\
    \textit{Rationale:} Assessing impact can help identify the most cited SESoS papers and understand how the research has been put forward.
\end{enumerate}

\subsection{Search strategy and eligibility criteria}
\label{subsec:search+selection}
Collecting all papers published in the SESoS editions from 2013 to 2023 followed a manual process. The second stage of the scoping review process (see \figurename~\ref{fig:process}) started by accessing the SESoS 2024 website\footnote{\url{http://sesos2024.icmc.usp.br}} and following the links to each workshop proceedings from 2013 to 2023 as the SESoS webpage is updated to list the proceedings of all the past editions. ACM and IEEE have published SESoS proceedings since the first edition of the workshop, so all the papers are available in the ACM Digital Library\footnote{\url{https://dl.acm.org}} and IEEE Xplore.\footnote{\url{https://ieeexplore.ieee.org/}} Next, all the published papers were downloaded from the proceedings of each workshop edition. The collection process resulted in 94 papers, 38 retrieved from the ACM Digital Library and 56 retrieved from IEEE Xplore.

Although SESoS has always focused on Software Engineering for SoS, not all papers appearing in the workshop strictly concern SoS since the topics of interest listed in the Calls for Papers are non-exhaustive and admit related themes. \added[id=EC]{For instance, software ecosystems are somewhat related to SoS due to their inherent distributed nature and the independence of their constituents, but these concepts are distinct from each other.\cite{Santos2024}} The presence of papers not directly focusing on SoS but notably on software ecosystems became evident mainly in the four joint editions of SESoS and WDES (2017, 2019, 2020, and 2021). Moreover, since its tenth edition (2022), the workshop changed its name, and the topics of interest listed in the current Call for Papers of SESoS now explicitly reflect addressing software ecosystems. \added[id=EC]{Therefore, the third stage of the scoping review process (see \figurename~\ref{fig:process}) adopted a criterion of excluding papers not explicitly related to SoS while screening all the retrieved SESoS papers.} This activity resulted in the exclusion of 37 papers, so 57 papers were included for analysis.

\subsection{Data extraction and analysis}
\label{subsec:dataextraction+analysis}
The data extraction step consisted of screening each paper to gather relevant data for answering the defined research questions. \tablename~\ref{tab:data-extraction} lists the data items extracted from SESoS papers, which were organized in an electronic spreadsheet.\footnote{\added[id=EC]{The interested reader can check the \hyperref[ex:data-availability]{Data Availability} section at the end of this article for information about the availability of the data extraction sheet.}} After filling out the data extraction spreadsheet, data items were quantitatively analyzed with the support of the R language and environment.\footnote{\added[id=EC]{The R scripts used for data analysis are publicly available online at \url{https://github.com/evertonrsc/analysis-sesos}.}} The \textit{Goals/Contributions} data extraction item underwent \added[id=EC]{content analysis}\cite{Elo2008} with inductive open coding procedures,\cite{Bandara2015,Stol2016} which allowed for classifying each paper into the SESoS topics of interest and the Software Engineering concepts from the 2012 ACM Computing Classification System (CCS)\cite{ACM2012} and identifying the research type it addressed.

\begin{center}
\begin{table*}
\caption{Data items extracted from SESoS papers.\label{tab:data-extraction}}
\begin{tabular*}{\textwidth}{@{}lp{12.6cm}@{}}
\toprule
Item & Description\\
\midrule
Year & SESoS edition where the paper was published\\
Title & Paper title\\
Paper type & Regular or short/position paper (regular papers: up to eight pages; short/position papers: up to four or five pages)\\
Citation count & Number of citations automatically retrieved from Google Scholar\\
Average citation count & Ratio between the total number of citations and the number of years since publication\\
Organic citation count & Number of citations excluding self-citations\\
Paper type & Regular or short/position paper\\
Authors' affiliation country & Countries of origin of the authors\\
SESoS topics & Categories of topics of interest as listed in the SESoS Call for Papers\\
Related Software Engineering field & One or more Level 2 concepts from the ACM CCS within the \textit{Software and its engineering category}\\
Goals/Contribution & Goals of the paper and its contributions\\
Research type & One of Validation, evaluation, solution proposal, philosophical, opinion, experience, secondary study\\
Empirical methods used (if applicable) & Empirical methods used in the research (notably in the evaluation research type)\\
SoS type(s) & Directed, acknowledged, collaborative, virtual SoS, or non-specified\\
Application domain or context & Application domain or context which the paper refers to, or non-specific\\
\bottomrule
\end{tabular*}
\end{table*}
\end{center}

The classification of SESoS papers comprised three tasks. First, the classification into the Software Engineering topics considered only the first two levels of the \textit{Software and its engineering} category of ACM CCS (see \tablename~\ref{tab:acm-ccs}) to reduce the probability of bias due to misclassification. Second, the classification into research types was done based on the categories presented in \tablename~\ref{tab:research-types}, which were adapted from Wieringa et al.'s work\cite{Wieringa2006} by including the \textit{secondary study} category due to the presence of this type of study in the analyzed papers. Third, the classification into the topics of interest of SESoS considered the ones presented in \tablename~\ref{tab:sesos-topics}, which were kept for most workshop editions. Although the division of the topics of interest into these categories has not been adopted since SESoS 2022, choosing to use them allowed for a uniform classification of the papers. \added[id=EC]{It is also important to mention that classifying papers into the SESoS topics of interest was not mutually exclusive, i.e., a given paper could span multiple categories.}

\begin{center}
\begin{table*}
\caption{Concepts within the \textit{Software and its engineering} of the 2012 ACM Computing Classification System\cite{ACM2012} used to classify SESoS papers.\label{tab:acm-ccs}}
\begin{tabular*}{\textwidth}{@{}>{\raggedright}p{3cm}p{14.2cm}@{}}
\toprule
Concept (Level 1) & Sub-concepts (Level 2)\\
\midrule
Software organization and properties & Contextual software domains, Software system structures, Software functional properties, Extra-functional properties\\
Software notations and tools & General programming languages, Formal language definitions, Compilers, Context-specific languages, System description languages, Development frameworks and environments, Software configuration management and version control systems, Software libraries and repositories, Software maintenance tools\\
Software creation and management & Designing software, Software development process management, Software development techniques, Software verification and validation, Software post-development issues, Collaboration in software development, Search-based software engineering\\ 
\bottomrule
\end{tabular*}
\end{table*}
\end{center}

\begin{center}
\begin{table*}
\caption{Taxonomy of research types -- adapted from Wieringa et al.\cite{Wieringa2006}\label{tab:research-types}}
\begin{tabular*}{\textwidth}{@{}>{\raggedright}lp{15cm}@{}}
\toprule
Research type & Description\\
\midrule
Validation & Investigation of the properties of a solution proposal through experiments, prototyping, formal analysis, and similar techniques.\\
Evaluation & Investigation of a problem or an implementation of a technique in practice. Causal relations and implications in terms of benefits and drawbacks are discussed upon solid empirical evidence.\\
Solution proposal & Proposal of a solution for a problem, either novel or a significant extension of an existing technique. The potential benefits and the applicability of the solution are demonstrated by a small example or proof-of-concept or via a thorough line of argumentation.\\
Philosophical & Sketch of a new way of looking at things, a new conceptual framework, etc.\\
Opinion & Expression of author's opinions whether a certain technique is good or bad, or how things should be done\\
Experience & Explanation of what and how something has been done in practice through case studies, surveys, and similar empirical research. It also encompasses the author's experience in a given context (e.g., an industrial case) or personal experience, typically describing insights and lessons learned.\\
Secondary study & Report on the conduction and results of a literature review (whether systematic or not) upon studies available in the literature.\\
\bottomrule
\end{tabular*}
\end{table*}
\end{center}

Finally, the scientometric analysis used co-author and citation analysis. The VOSViewer software tool\footnote{\url{https://www.vosviewer.com}} was used to map and visualize research collaborations in SESoS papers, thus allowing for deductions about people and locations carrying out research in Software Engineering for SoS. 
\blue{Moreover, the citation counts of each SESoS paper were automatically scraped from Google Scholar by a program implemented in the Go programming language\footnote{\added[id=EC]{\url{https://github.com/evertonrsc/citations-gscholar}}} powered by the SerpAPI Google Scholar API.\footnote{\url{https://serpapi.com/google-scholar-api}}}

\section{Results}
\label{sec:results}
This section presents the results of this study. Section~\ref{subsec:distribution-studies} provides an overview of the distribution of the selected SESoS papers. Sections~\ref{subsec:rq1} to~\ref{subsec:rq7}) provide answers to the defined RQs based on the information extracted from those studies. \added[id=EC]{Section~\ref{subsec:threats-to-validity} discusses some possible threats to the validity of the study reported in this article.}

\subsection{Distribution of studies}
\label{subsec:distribution-studies}
\begin{figure*}
\centerline{\includegraphics[scale=0.14]{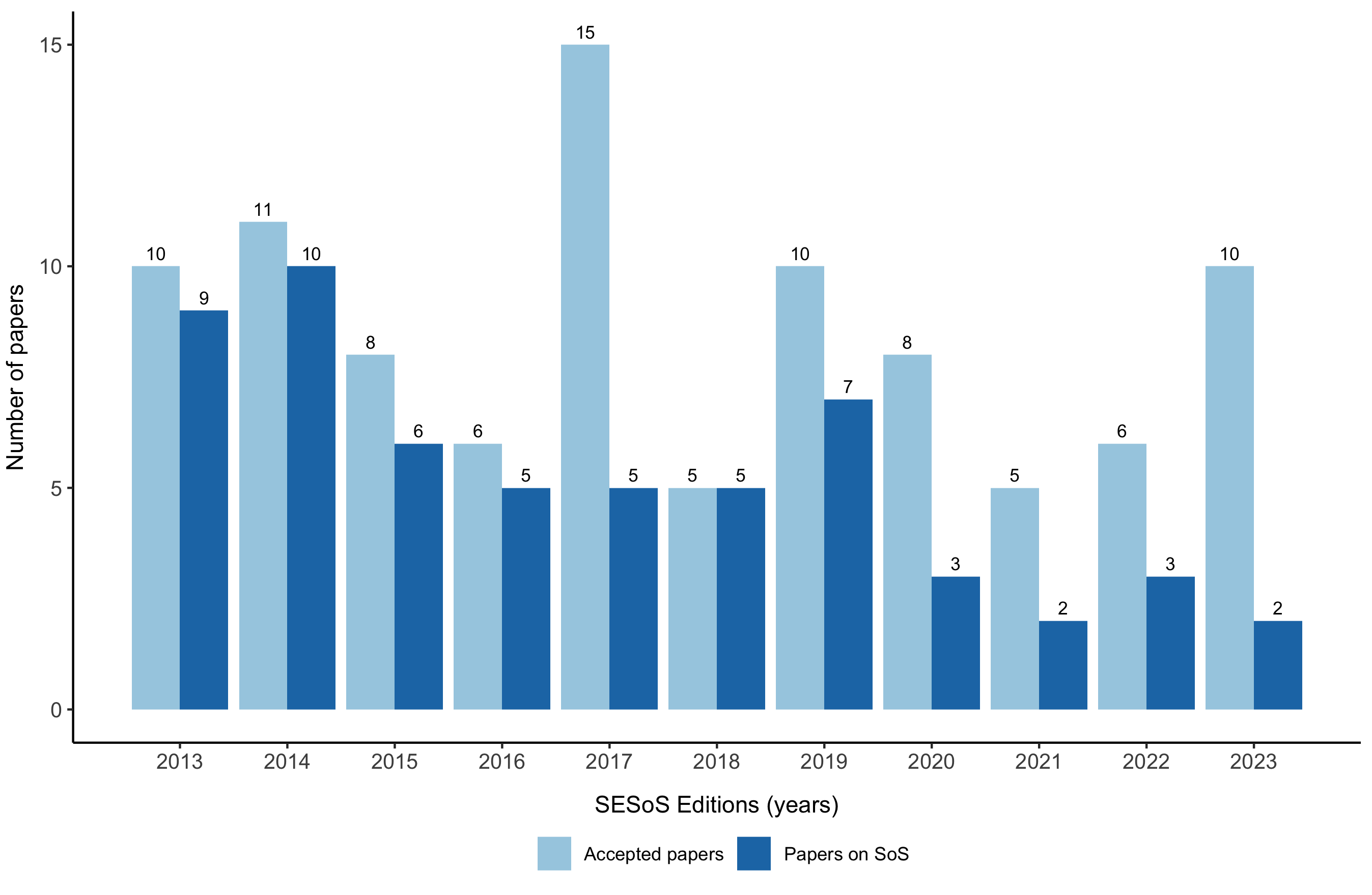}}
\caption{Temporal distribution of papers along the SESoS editions.\label{fig:numpapers}}
\end{figure*}

\figurename~\ref{fig:numpapers} depicts the temporal distribution of papers along the SESoS editions, with an average of 5.18 papers explicitly related to SoS per year. The chart also shows a significant decrease in the number of papers in 2021 and 2022, undoubtedly due to the COVID-19 pandemic, which impacted scientific production worldwide.\cite{Raynaud2021,Heo2022} However, the pandemic could not have affected the number of papers in 2020 because the submission deadline to the SESoS edition in that year was in mid-January, before the COVID-19 outbreak and the first acute phase of the crisis.

\added[id=EC]{The chart in \figurename~\ref{fig:numpapers} shows a peak regarding the number of submitted papers in 2017. In that year, SESoS was held as a joint workshop with WDES, thereby broadening the topics of interest towards SoS, software ecosystems, and distributed software development so that the amplification of the workshop's scope may have led to more submissions. A higher number of submitted papers was also observed in 2019 and 2020 for the second and third editions of the joint SESoS/WDES workshop compared to previous years. On the other hand, it is also possible to observe a decline in the number of papers in SESoS 2021 and 2022, which may be a consequence of the COVID-19 pandemic. The number of accepted papers increased in SESoS 2023 as an indication of a recovery in the following years, but this is a conjecture to be confirmed with data from the forthcoming workshop editions.}

Another important facet to analyze is the proportion of papers on SoS to the number of papers accepted in SESoS editions. The topics of interest of the first four editions (2013-2016) and the sixth one (2018) significantly focused on Software Engineering for software-intensive SoS, thus resulting in the majority of the papers accepted to the workshop being related to this theme, ranging from 75\% (2015) to 100\% (2018). In 2017 and from 2019 to 2021, SESoS was held as a joint workshop with WDES. Consequently, the topics of interest included other themes not necessarily related to SoS, such as software ecosystems and distributed software development. Moreover, the last two editions of SESoS (2022 and 2023) have explicitly incorporated other topics beyond SoS, mainly those related to software ecosystems. These facts also explain the reduced number of papers about SoS in the SESoS workshop compared to the total of accepted ones.

\begin{mdframed}[innerleftmargin=0.25cm,innerrightmargin=0.25cm,innertopmargin=0.25cm,innerbottommargin=0.25cm,skipabove=0.5cm,linecolor=blue]
\added[id=EC]{Summary of findings:}
\begin{itemize}[leftmargin=*,topsep=-4pt]
	\item \added[id=EC]{An average of 5.18 papers on SoS have been published in SESoS between 2013 and 2023.}
	\item \added[id=EC]{The COVID-19 pandemic affected the number of papers published in SESoS 2021 and 2022.}
	\item \added[id=EC]{The number of papers explicitly focused on SoS has decreased in recent years with the joint editions of the SESoS and WDES workshops and the amplification of the topics of interest for SESoS to incorporate software ecosystems and related themes.}
\end{itemize}
\end{mdframed}

\subsection{Research loci \blue{(RQ1)}}
\label{subsec:rq1}
RQ1 concerns identifying where the research on Software Engineering for SoS has been conducted, including collaboration networks and industry involvement. \tablename~\ref{tab:papers-countries} shows the number of papers per affiliation country of the authors. It is possible to notice that the research has been concentrated in Brazil, Europe (more specifically, France, Spain, Germany, Sweden, Austria, and Italy), and South Korea.

\begin{center}
\begin{table*}
\caption{Contributing countries to SESoS between 2013 and 2023.\label{tab:papers-countries}}
\begin{tabular*}{\textwidth}{@{}l@{\hspace*{8cm}}l@{}}
\toprule
Country & Number of papers\\
\midrule
Brazil & 25 (47.83\%)\\
France & 14 (24.56\%)\\
Sweden & 6 (10.52\%)\\
South Korea & 6 (10.52\%)\\
Germany & 6 (10.52\%)\\
Spain & 5 (8.77\%)\\
Italy & 4 (7.02\%)\\
USA & 4 (7.02\%)\\
Austria & 2 (3.51\%)\\
United Kingdom & 1 (1.75\%)\\
The Netherlands & 1 (1.75\%)\\
Tunisia & 1 (1.75\%)\\
Australia & 1 (1.75\%)\\
\bottomrule
\end{tabular*}
\end{table*}
\end{center}

\begin{figure*}
\centerline{\includegraphics[scale=0.2]{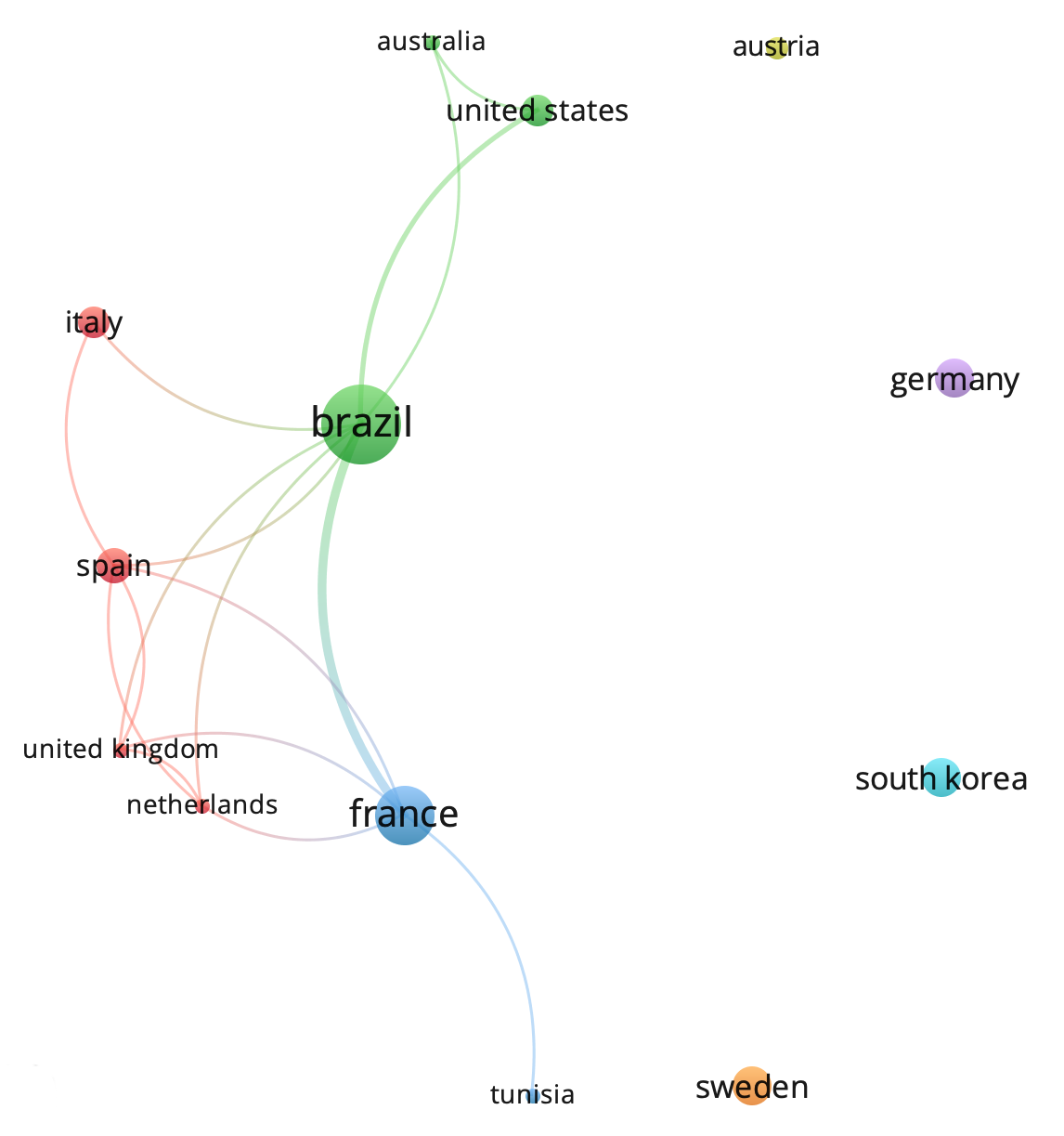}}
\caption{Co-authorship country network for SESoS papers (2013-2023).\label{fig:coauthorship-countries}}
\end{figure*}

\figurename~\ref{fig:coauthorship-countries} shows a co-authorship network based on the authors' affiliation countries. In this network, the larger the node, the greater the number of papers; the thicker the link, the greater the number of co-authored papers. An analysis of the co-authorship network reveals four loosely-tied, small clusters, thereby indicating that research is fragmented, with few international collaborations. Brazil and France are the countries with the two highest number of papers in SESoS (25 and 14, respectively) and the most robust collaborations (six co-authored papers).

The analysis of the authors' affiliation also concerned the involvement of industry and collaborations with academia. Almost all the papers are authored by researchers from universities and research institutes, and only one has an author from industry (paper P9). Nonetheless, it is worthwhile mentioning that several papers report research in collaboration with or applied to industrial scenarios, such as papers P43, P45, P56, and P57. Some institutions contributing to SESoS are well-known for significant applied research in partnership with industry, such as the Fraunhofer Institute for Experimental Software Engineering in Kaiserslautern, Germany. \added[id=EC]{Even though SESoS has consolidated primarily as an academic forum, it undoubtedly should use strategies to attract more contributions and participants from the industry in future workshop editions.}

\begin{mdframed}[innerleftmargin=0.25cm,innerrightmargin=0.25cm,innertopmargin=0.25cm,innerbottommargin=0.25cm,skipabove=0.5cm,linecolor=blue]
\added[id=EC]{Summary of findings:}
\begin{itemize}[leftmargin=*,topsep=-4pt]
	\item \added[id=EC]{Research reported in SESoS between 2013 and 2023 is mainly concentrated on Brazil and France. There are also noticeable contributions from Sweden, South Korea, and Germany.}
	\item \added[id=EC]{The research landscape is fragmented, with few international collaborations. The most robust collaborations are observed between Brazil and France.}
	\item \added[id=EC]{The involvement of the industry in the research is tiny, even though some papers report research applied to industrial scenarios.}
\end{itemize}
\end{mdframed}

\subsection{Prevailing topics \blue{(RQ2)}}
\label{subsec:rq2}
\begin{figure*}
\centerline{\includegraphics[scale=0.5]{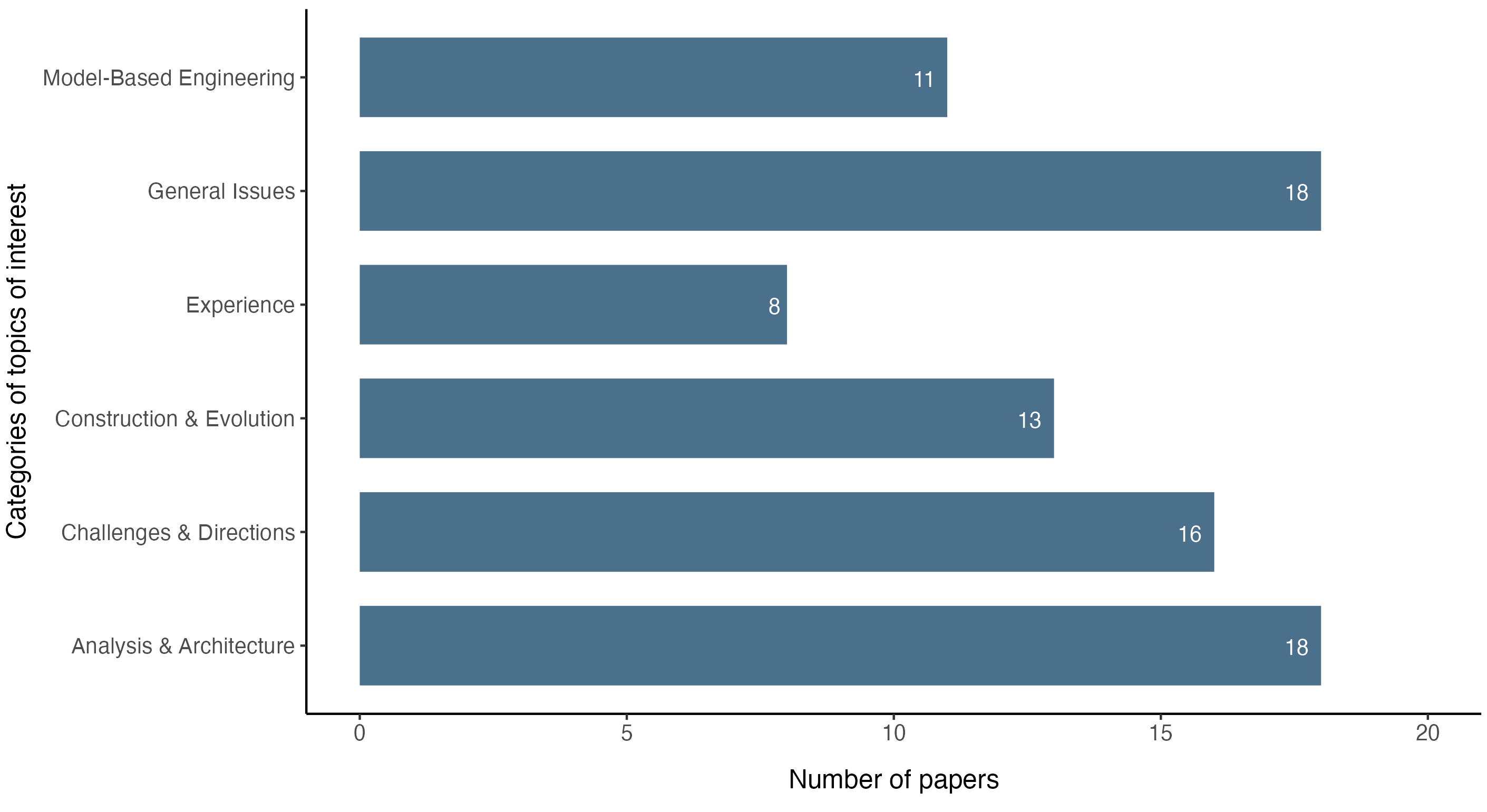}}
\caption{Categories of topics of interest addressed by SESoS papers.\label{fig:topics-papers}}
\end{figure*}

RQ2 aims to classify the papers across the topics presented in the SESoS Calls for Papers and understand how the interest in those topics changed over time. \figurename~\ref{fig:topics-papers} shows the classification of the papers according to the categories of topics of interest in SESoS (see \tablename~\ref{tab:sesos-topics}). It is possible to notice that one of the most addressed categories is \textit{Analysis \& Architecture} (18 papers), which comprises the design, representation, evaluation, and evolution of SoS software architectures. In both traditional systems and SoS, software architectures are key elements for determining the success of these systems, but novel architectural solutions are required to handle the inherent complexity, evolutionary development, and emergent behavior of SoS.\cite{Oquendo2016a} ISO/IEC/IEEE 42010:2022,\cite{ISO42010} one of the main international standards related to software architectures, indeed corroborates the importance of software architectures for SoS. Nonetheless, the literature does not sufficiently discuss architectural activities such as evaluation, implementation, and evolution in the context of SoS.\cite{Cadavid2020}

\textit{General Issues} is also the most addressed category of topics of interest (18 papers). Data collected from SESoS papers reveal a significant interest in quality issues for SoS, focusing on the interoperability among constituent systems, which is fundamental for SoS to be built and work. In SESoS 2015, paper P21 reported the results of an SLR on quality attributes in the SoS context, pointing out interoperability as one of the five most relevant quality attributes. In the following year, paper P30 presented the results of another SLR focused on integrating constituent systems in SoS, highlighting that this is a high-demand concern. In general, the literature on architecting SoS has mainly focused on quality attributes concerning how constituent systems cooperate with each other.

Another recurrent topic of interest concerns SoS VV\&T, which appears in 12 papers. These processes are of paramount importance considering the distinguishing characteristics of SoS and that many of these systems are designed for mission-critical domains, thus requiring robust techniques to ensure their proper operation. In SESoS 2018, paper P39 brought interesting insights into how the SoS characteristics impact VV\&T processes, which are well-established in Software Engineering but are limited or cannot be applied to SoS.

\figurename~\ref{fig:topics-papers} also shows that more than a quarter of SESoS papers on SoS has accounted for the \textit{Challenges \& Directions} category ($n = 16$), which is related to raising and discussing challenges and future perspectives of research on Software Engineering for SoS. This characteristic of the papers aligns with the purposes of the SESoS workshop, which aims to be a forum where researchers and practitioners can discuss ideas and experiences, analyze research and development issues, and propose inspiring visions. Secondary studies play a relevant role in achieving these purposes, and nine papers (P8, P17, P18, P19, P21, P30, P33, P46, and P49) reported the results of SLRs and SMS addressing different topics discussing relevant gaps in research and development related to SoS.

\begin{figure*}
\centerline{\includegraphics[scale=0.65]{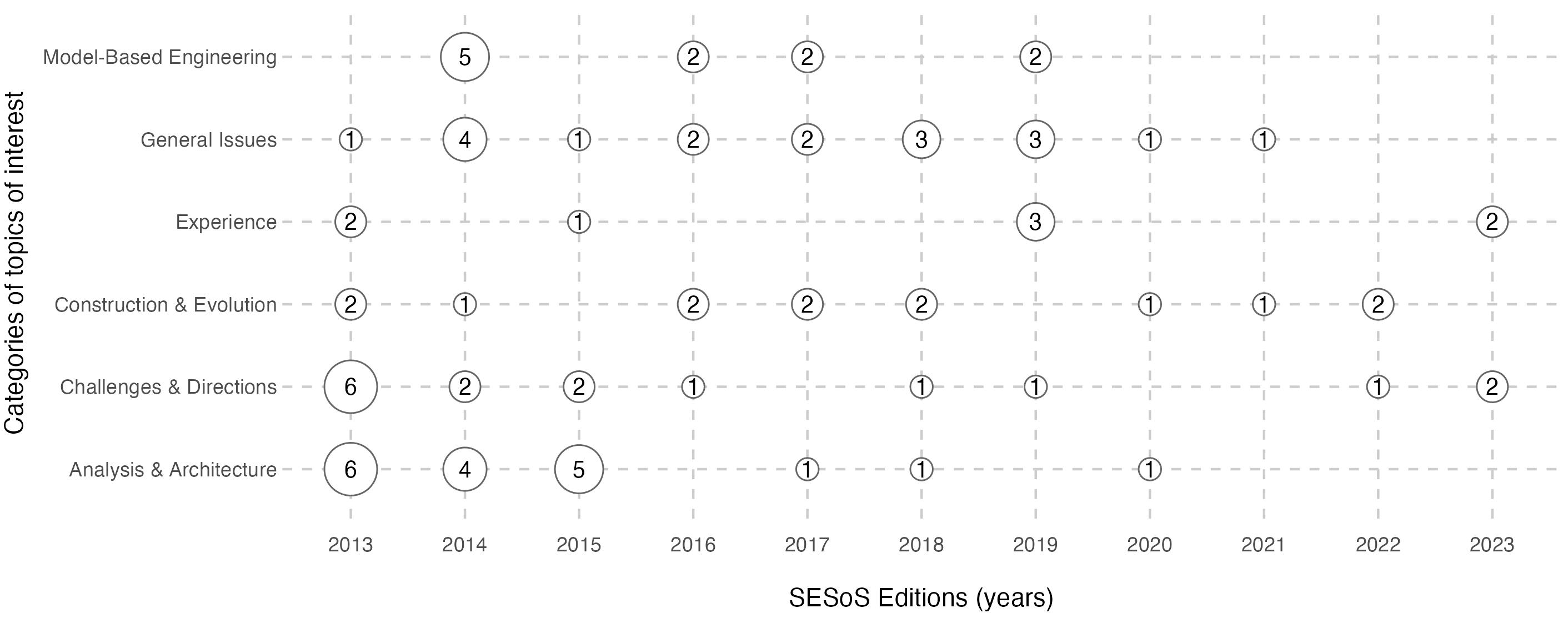}}
\caption{Distribution of the topics addressed by SESoS papers over the years.\label{fig:topics-years}}
\end{figure*}

\figurename~\ref{fig:topics-years} shows the temporal distribution of the topics addressed by SESoS papers. It is possible to observe that the topics within the \textit{Analysis \& Architecture} category stood out in the first editions of the workshop, possibly because (i) they were held with ECSA, one of the main international conferences on Software Architecture field, and (ii) most of the authors of those papers are recognized for their research related to software architectures, so driving the research towards SoS software architectures would be a natural choice. The noteworthy number of papers in the \textit{Challenges \& Directions} category is also observed in the first edition of the workshop (2013), which was significantly characterized by papers discussing different challenges and research perspectives on exploring Software Engineering for software-intensive SoS. The \textit{Construction \& Evolution} and \textit{General Issues} categories remained relatively constant over the years, with papers focusing on different concerns related to building effective, high-quality SoS.

\begin{figure*}
\centerline{\includegraphics[scale=0.65]{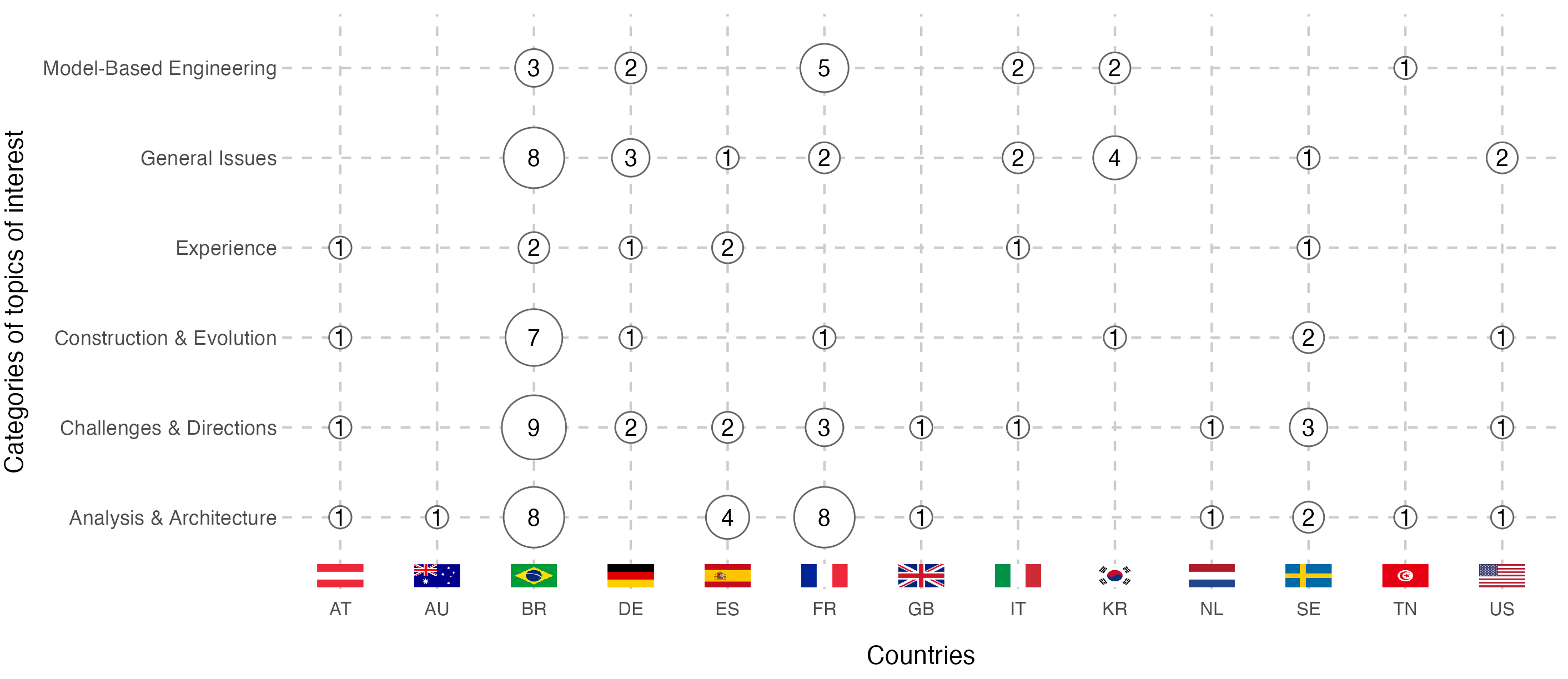}}
\caption{Distribution of the topics addressed by SESoS papers based on country of authorship (represented by an alpha-2 code as in ISO 3166).\label{fig:topics-countries}}
\end{figure*}

\added[id=EC]{\figurename~\ref{fig:topics-countries} shows the distribution of the topics of interest covered by SESoS papers over location based on authors' countries. Researchers from Brazil and France, the countries with the most co-authored papers in SESoS from 2013 to 2023 (see Section~\ref{subsec:rq1}), have more contributions on topics related to SoS software architectures. Such an interest can be explained by the fact that these countries have relevant research groups actively contributing to the international Software Architecture community for over a decade. Furthermore, the interest of researchers from Brazil who have published in SESoS in the \textit{Challenges \& Directions} category of topics of interest is remarkable. As a matter of fact, all the secondary studies, i.e., SLRs and SMS, published in SESoS from 2013 to 2023 ($n = 9$) were authored by researchers from Brazil.}

\added[id=EC]{\figurename~\ref{fig:topics-countries} also shows the focus of papers authored by researchers from South Korea on general issues related to Software Engineering for SoS, mainly the verification and validation of built SoS (papers P26, P34, P42, and P47). While researchers from France and Germany are also interested in this topic, there are no collaborations among these countries, as shown in \figurename~\ref{fig:coauthorship-countries}. Similar behavior is noticed for the \textit{Model-Based Engineering} category, for which 11 papers are authored from six different countries, but only two papers account for collaborations between Brazil and France (P18 and P35), and one between Tunisia and France (P12). These observations lead to the conclusion that some research groups within the SESoS community sometimes work on the same or similar subjects but do not collaborate. The SESoS workshop shall take further action toward accomplishing one of its missions of paving the way for a more structured community effort.}


\begin{mdframed}[innerleftmargin=0.25cm,innerrightmargin=0.25cm,innertopmargin=0.25cm,innerbottommargin=0.25cm,skipabove=0.5cm,linecolor=blue]
\added[id=EC]{Summary of findings:}
\begin{itemize}[leftmargin=*,topsep=-4pt]
	\item \added[id=EC]{Most SESoS papers have addressed SoS software architectures, focusing more on design and representation and less on evaluation and evolution.}
	\item \added[id=EC]{Interoperability among constituent systems and SoS VV\&T are critical concerns addressed in the research reported in SESoS papers.}
	\item \added[id=EC]{Secondary studies such as SLRs and SMS often appear in SESoS to raise and discuss challenges and open issues related to the research on Software Engineering for SoS.}
	\item \added[id=EC]{Research groups from different locations sometimes work on the same or similar topics of interest but do not collaborate.}
\end{itemize}
\end{mdframed}

\subsection{Software Engineering topics \blue{(RQ3)}}
\label{subsec:rq3}
\begin{figure*}
\centerline{\includegraphics[scale=0.16]{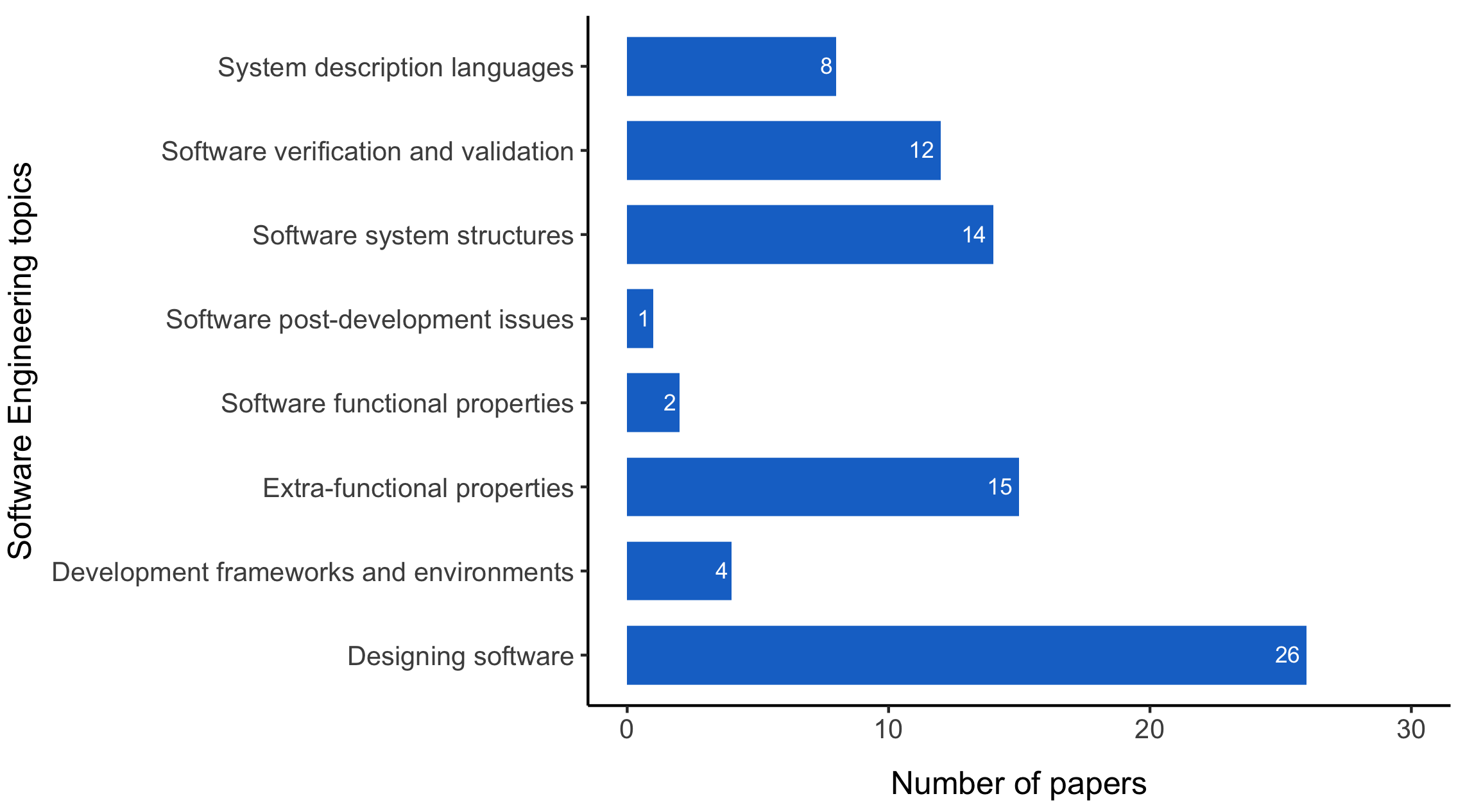}}
\caption{Classification of SESoS papers according to the \textit{Software and its engineering} category of the 2012 ACM Computing Classification System.\label{fig:se-topics}}
\end{figure*}

RQ3 seeks to cross-reference the topics addressed in SESoS papers and the main topics in the Software Engineering field as organized by the ACM CCS under the \textit{Software and its engineering} category (see Table~\ref{tab:acm-ccs}). \figurename~\ref{fig:se-topics} shows the classification of the papers according to the concepts defined in the ACM CCS taxonomy. \textit{Designing software} is the topic with the highest occurrence, accounting for almost half of SESoS papers on SoS ($\approx$ 45.6\%). This prevalence might be due to research reported in those papers seeking to propose approaches, methods, and techniques to construct SoS.

\figurename~\ref{fig:se-topics} also shows results in line with observations of the topics of interest of the SESoS workshop (see Section~\ref{subsec:rq3}). \textit{Extra-functional properties} represents a topic of significant interest (15/57 papers) with emphasis on interoperability, followed by \textit{Software system structures} (14/57 papers) mainly due to the focus on the design of SoS software architectures. \textit{Software verification and validation} is another relevant topic (12/57 papers) considering the critical role of VV\&T in many SoS.

It is also possible to notice that only one paper addressed the \textit{Software post-development issues} topic, which comprises software evolution. Evolutionary development, one of the distinguishing (and complicating) characteristics of SoS, allows these systems to continuously evolve to respond to changes, whether related to the environment, their constituent systems, their functionalities or missions, or the independent evolution of constituent systems. This low number of papers concerning SoS evolution might indicate that research on Software Engineering for SoS has not sufficiently focused on how SoS and their constituent systems evolve, thereby revealing a gap in this context.

\begin{mdframed}[innerleftmargin=0.25cm,innerrightmargin=0.25cm,innertopmargin=0.25cm,innerbottommargin=0.25cm,skipabove=0.5cm,linecolor=blue]
\added[id=EC]{Summary of findings:}
\begin{itemize}[leftmargin=*,topsep=-4pt]
	\item \added[id=EC]{Most SESoS papers propose approaches, methods, and techniques for designing SoS, significantly focusing on software architectures.}
	\item \added[id=EC]{Extra-functional properties, such as interoperability, are among the leading interests of research reported in SESoS.}
	\item \added[id=EC]{VV\&T is a relevant research topic for SESoS papers.}
	\item \added[id=EC]{The number of papers concerning SoS post-development issues, including evolutionary development, is significantly low.}
\end{itemize}
\end{mdframed}

\subsection{Research types \blue{(RQ4)}}
\label{subsec:rq4}
\begin{figure*}
\centerline{\includegraphics[scale=0.55]{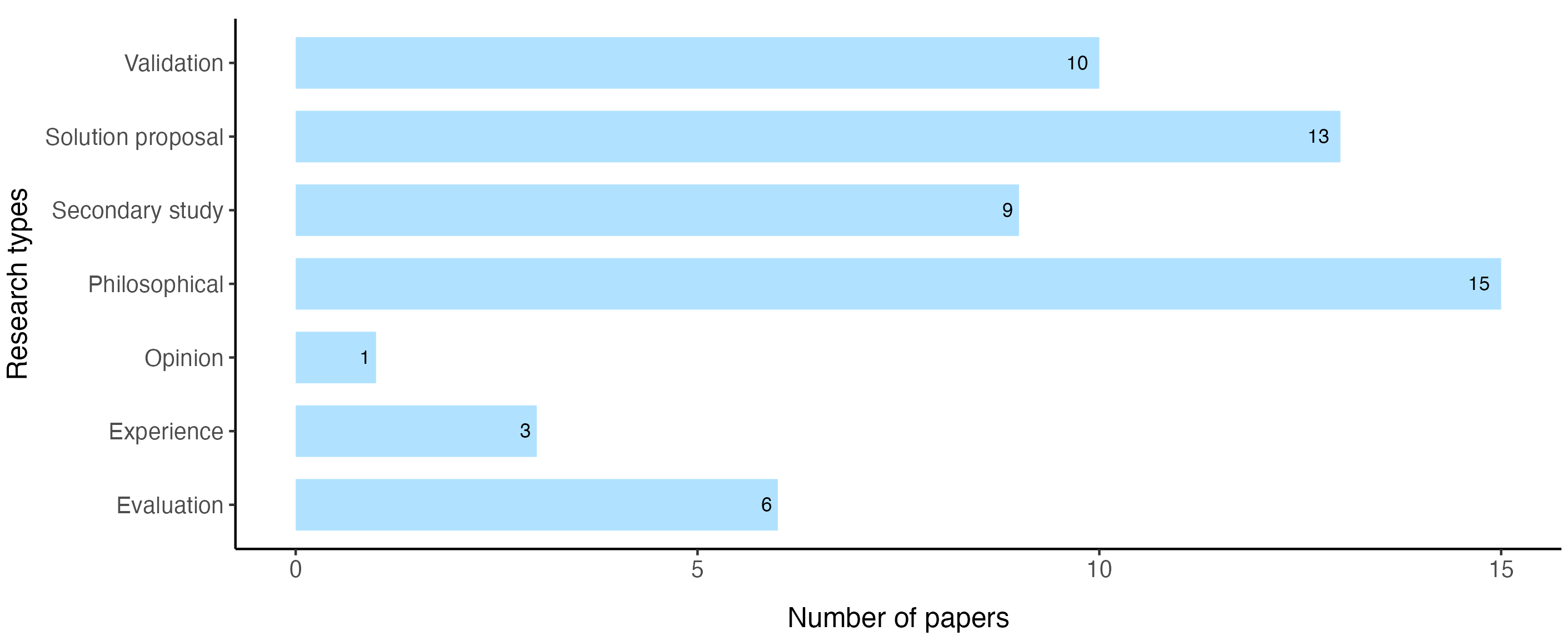}}
\caption{Classification of SESoS papers into research types.\label{fig:research-types}}
\end{figure*}

RQ4 aims to analyze the research types addressed by SESoS papers from 2013 to 2023, also as an indicator of the nature and maturity of the contributions. \figurename~\ref{fig:research-types} shows the classification of SESoS papers according to the type of research as summarized in \tablename~\ref{tab:research-types}. More than a quarter of the papers ($n = 15$) are \textit{philosophical} papers, i.e., they discuss current issues and promising research directions for SoS from a Software Engineering perspective. This type of research occurred more frequently (five out of nine papers) in the first edition of the workshop when the aim was to set up an agenda and lay the foundations for future developments in Software Engineering for SoS. Furthermore, the last two editions of the workshop brought papers on SoS focusing on the discussion of challenges related to SoS considering scenarios that have been drawing attention from academia and industry in recent years, such as digital twins (paper P55 in SESoS 2022) and cyber-physical systems (paper P57 in SESoS 2023). In SESoS 2023, paper P56 aimed to discuss the dynamic nature of SoS, a topic that still needs to be more explored in the literature, even though it is inherent to this class of systems.

The second most addressed research type in SESoS papers from 2013 to 2023 is \textit{solution proposal} ($\approx$ 22.8\%). In general, these papers present the proposal of a solution to a relevant research problem without going any further in validating or evaluating it but rather presenting a demonstration of the feasibility of their proposal through a proof of concept or example. These data indicate a lower maturity of some research presented in the workshop, considering the small number of papers with more robust evaluation ($n = 6$) primarily using computational experiments as an empirical method (four papers). Nevertheless, it is essential to highlight that the primary purpose of any workshop is to provide an opportunity for researchers to exchange and discuss scientific and engineering ideas at an early stage before they have matured to conference or journal publications. \added[id=EC]{The low involvement of the industry, already observed in Section~\ref{subsec:rq1}, can also be related to having more solution proposals and less direct applications to problems identified in practice.}

Another research type often appearing in SESoS papers is secondary studies (nine out of 57 papers), specifically reporting SLR results. This research type is of significant importance, considering that the typical outcomes of SLRs are (i) the presentation of a comprehensive overview of the investigated topic and (ii) the identification and discussion of gaps found in the literature to commission future research.\cite{Petersen2008,Petersen2015} The secondary studies presented in SESoS from 2013 to 2023 cover diverse topics, mainly designing and constructing SoS (papers P8, P18, P19, P30, P33). In SESoS 2019, paper P48 shed light on a specific class of SoS, the so-called systems of information systems (SoIS). SoIS comprise operationally independent information systems that interoperate with other constituents to provide unique capabilities.\cite{Saleh2015,Fernandes2019} Besides inheriting the characteristics of SoS, SoIS present specific ones, such as (i) the existence of information flows among the constituent information systems, (ii) a business-oriented process nature, and (iii) the interoperability among the constituent information systems and their organizations generating information and adding value.\cite{Fernandes2022} These additional, specific characteristics pose new challenges for research and development in information systems.

\begin{figure*}
\centerline{\includegraphics[scale=0.6]{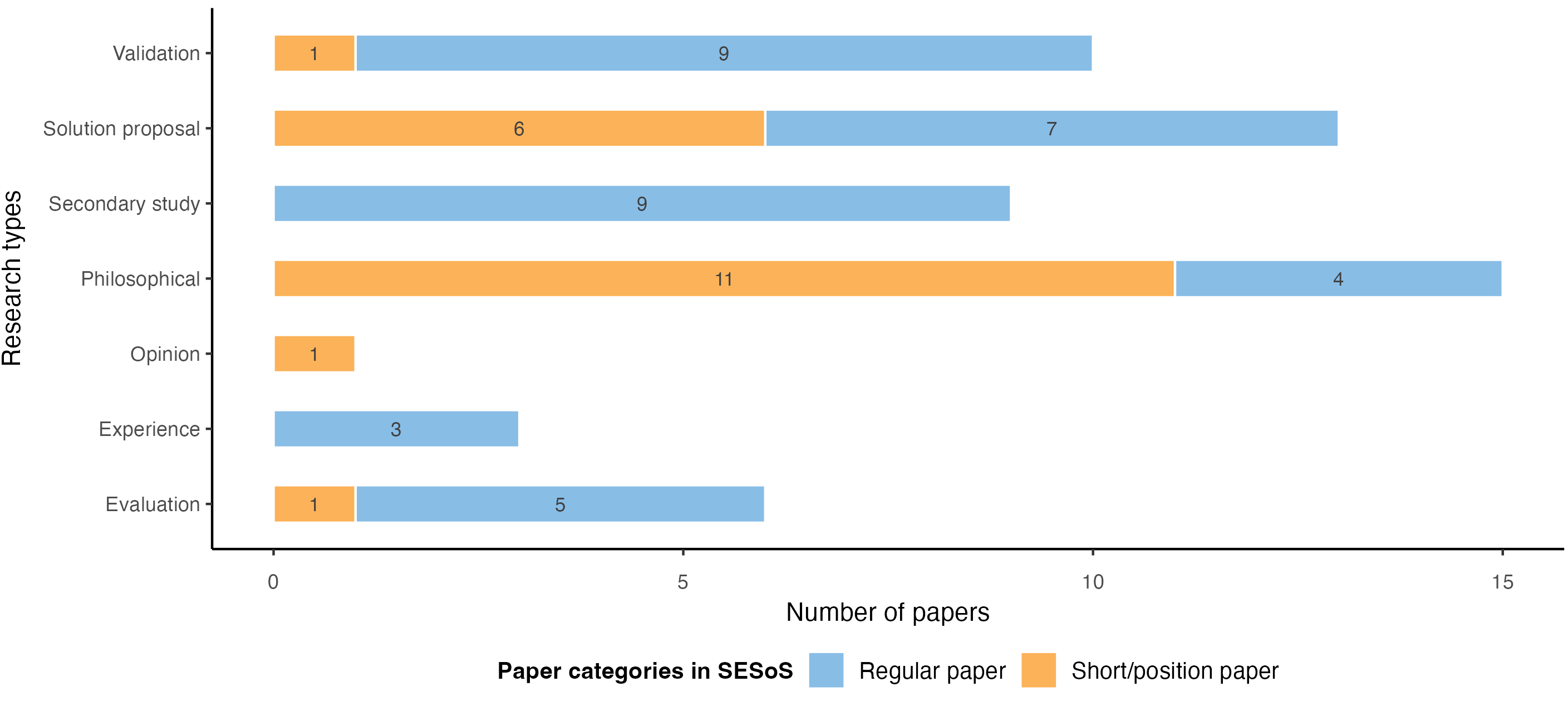}}
\caption{Classification of SESoS papers into research types and paper categories.\label{fig:research-types_vs_papers}}
\end{figure*}

\figurename~\ref{fig:research-types_vs_papers} depicts some relationships between research types and the categories of papers solicited by SESoS, i.e., regular or short/position papers. Most SESoS papers from 2013 to 2023 are regular ones ($\approx$ 64.9\%), indicating that the papers generally present substantial content. Papers addressing \textit{validation}, \textit{evaluation}, \textit{experience}, and \textit{secondary study} ($n = 26$) present the study results or the reported experiences in more detail, thereby falling into the category of regular papers. On the other hand, it is possible to notice that SESoS papers addressing \textit{solution proposal} ($n = 13$) vary in different degrees so that regular papers addressing this research type tend to present more mature, detailed solutions than short/position papers.

\begin{mdframed}[innerleftmargin=0.25cm,innerrightmargin=0.25cm,innertopmargin=0.25cm,innerbottommargin=0.25cm,skipabove=0.5cm,linecolor=blue]
\added[id=EC]{Summary of findings:}
\begin{itemize}[leftmargin=*,topsep=-4pt]
	\item \added[id=EC]{Many papers in SESoS from 2013 to 2023 have discussed current issues and future research directions on Software Engineering for SoS.}
	\item \added[id=EC]{Many papers in SESoS have brought contributions as low-maturity solution proposals.}
	\item \added[id=EC]{A few SESoS papers report empirical studies with a more robust evaluation.}
	\item \added[id=EC]{Secondary studies reporting results of SLRs and SMS appear often in SESoS.}
	\item \added[id=EC]{Most of the contributions to SESoS are regular papers.}
\end{itemize}
\end{mdframed}

\subsection{Application domains or contexts \blue{(RQ5)}} 
\label{subsec:rq5}
\begin{figure*}
\centerline{\includegraphics[scale=0.55]{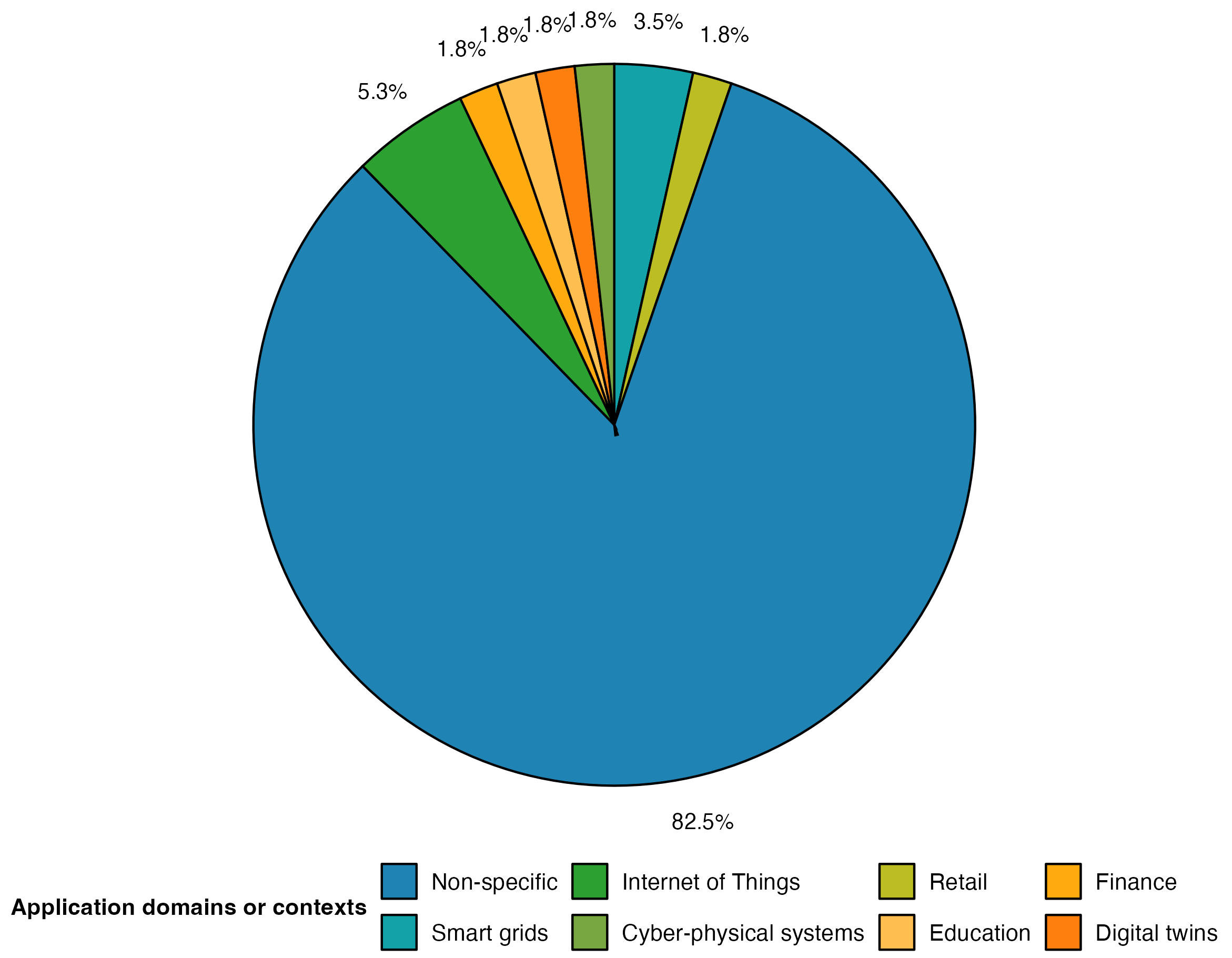}}
\caption{Application domains and contexts of papers in SESoS 2013--2023.\label{fig:domains}}
\end{figure*}

RQ5 seeks to identify the application domains or contexts SESoS papers address to assess how specific the reported contributions are. \figurename~\ref{fig:domains} shows that most papers (82.5\%) do not concern a specific domain, i.e., their contributions are somewhat generic. Nonetheless, some papers demonstrate their solution proposals, validation, and evaluation in domains such as industry automation (paper P2), industrial plants (paper P9), military (papers P10, P12, P37), and disaster prevention and management (papers P23, P26, P39, P42, P47). On the other hand, some papers report SoS in other contexts, such as the Internet of Things (papers P6, P34, and P51) and cyber-physical systems (paper P7), even though the research is not specific to a given application domain for systems in these contexts.

\figurename~\ref{fig:domains} also shows other application domains with minor occurrences within the SESoS papers. Smart grids ($n = 2$), retail, education, and finance ($n = 1$ each) are domains addressed by some papers. Papers P3 and P24 describe the proposal to design a reference architecture for large-scale smart grid SoS. Papers P43 and P45 respectively report the experiences of SoS in retail and finance.

\added[id=EC]{The collected data allowed for outlining two insights. First, research in Software Engineering for SoS shall be better balanced with practice to allow for a more concrete realization of the proposed approaches and solutions.} \added[id=FO]{Second, the SESoS workshop could evolve to include a new type of submission that would be more convenient to attract participants from the industry than the current kinds of contributions (regular and position/short papers), which are mainly used by academics. This would enlarge the workshop's scope to better consider industrial problems in the research on Software Engineering for SoS.}

\begin{mdframed}[innerleftmargin=0.25cm,innerrightmargin=0.25cm,innertopmargin=0.25cm,innerbottommargin=0.25cm,skipabove=0.5cm,linecolor=blue]
\added[id=EC]{Summary of findings:}
\begin{itemize}[leftmargin=*,topsep=-4pt]
	\item \added[id=EC]{Most SESoS papers do not concern any specific domain.}
	\item \added[id=EC]{Research in contexts such as the Internet of Things and cyber-physical systems has appeared in SESoS papers from 2013 to 2023 despite not being related to specific domains.}
	\item \added[id=EC]{There is a need for more concretization of solution proposals in practice.}
\end{itemize}
\end{mdframed}

\subsection{\blue{SoS types (RQ6)}}
\label{subsec:rq6}
\added[id=EC]{From the work of Maier\cite{Maier1998} and Dahmann and Baldwin,\cite{Dahmann2008} the ISO/IEC/IEEE 21841 International Standard\cite{ISO21841} defined a taxonomy to classify SoS into four types according to the degree of governance over constituent systems. \textit{Directed SoS} are SoS created and managed to fulfill specific purposes and whose constituent systems are subordinated to a central control at the SoS level, even though they maintain their operational and managerial independence. With less control, \textit{acknowledged SoS} are SoS with recognized objectives, a designated manager, and resources, although constituent systems retain their independence and changes on them rely on agreements with the SoS. \textit{Collaborative SoS} do not have centralized management, and constituent systems interact more or less voluntarily (i.e., they may decide to provide or deny service) to fulfill agreed missions at the SoS level. In \textit{virtual SoS}, constituent systems self-organize without a centrally managed entity and agreed purposes. RQ6 aims to identify the SoS type(s) concerned by SESoS papers to understand the extent of research they report since the SoS type can directly influence the nature of the contributions.}

\added[id=EC]{The identification of the SoS types considered only studies describing an SoS to which the solutions or approaches have been proposed or applied ($n$ = 32). Therefore, studies that have been classified as philosophical papers, opinion, and secondary studies (see Section~\ref{subsec:rq4}) were left out of this analysis as they do not concern any concrete SoS. Moreover, some of those studies explicitly state the SoS type; in others, it could be inferred from the description of the presented case of SoS. When solution proposals or approaches presented in the studies did not concern a given SoS type, or there was not enough evidence to infer it, it has been assumed that the SoS type has not been specified.}

\begin{figure*}
\centerline{\includegraphics[scale=0.5]{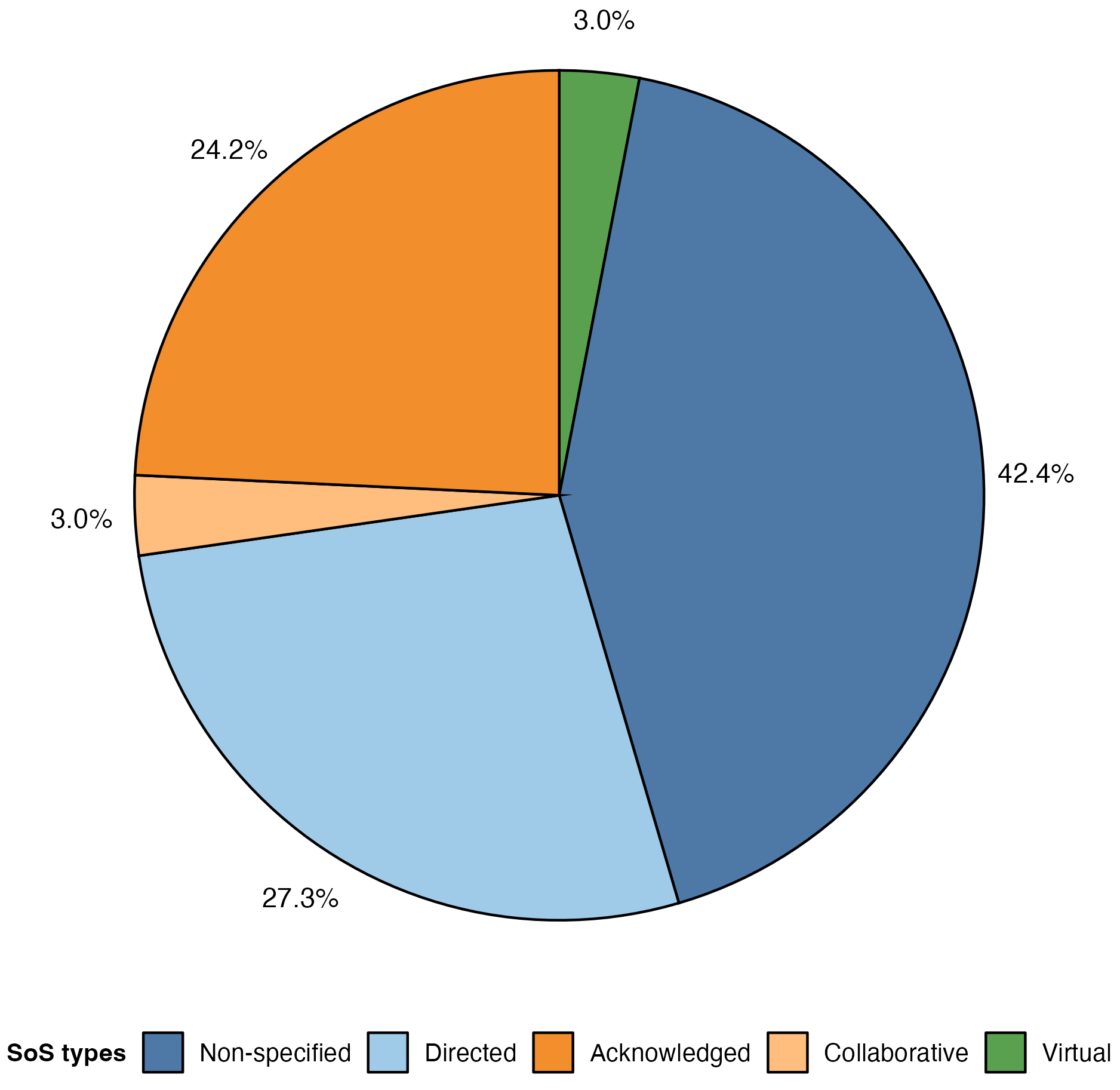}}
\caption{SoS types concerned by research reported in SESoS 2013-2023.\label{fig:sostypes}}
\end{figure*}

\added[id=EC]{\figurename~\ref{fig:sostypes} shows that the SoS type is not specified in almost half of the SESoS papers classified as solution proposals, experience reports, validation, or evaluation (14/32). On the other hand, nearly a quarter of the studies concerned directed SoS (9/32), and another quarter acknowledged SoS (8/32). This prevalence of directed and acknowledged SoS leads to some insights. First, governance mechanisms in collaborative and virtual SoS are less evident and need to be better understood, while directed and acknowledged SoS seem more tractable due to the existence of a central authority that manages the SoS to a greater or smaller degree and aligns missions to be accomplished. Second, to some extent, directed and acknowledged SoS still admit applying or adapting existing approaches that system and software engineers already know.\cite{Henshaw2023}}

\begin{mdframed}[innerleftmargin=0.25cm,innerrightmargin=0.25cm,innertopmargin=0.25cm,innerbottommargin=0.25cm,skipabove=0.5cm,linecolor=blue]
\added[id=EC]{Summary of findings:}
\begin{itemize}[leftmargin=*,topsep=-4pt]
	\item \added[id=EC]{Most of the SESoS papers proposing solutions or applying approaches do not specify the type of SoS they are concerned with.}
	\item \added[id=EC]{Many studies appearing in the SESoS workshop from 2013 to 2023 address directed and acknowledged SoS.}
\end{itemize}
\end{mdframed}

\subsection{Research impact \blue{(RQ7)}}
\label{subsec:rq7}
RQ7 concerns assessing the impact of SESoS papers by analyzing how much they have been cited. Citation count is the most straightforward metric for measuring scientific impact despite its criticism and many factors that may affect it, such as motivations for citing a paper and the influence of publication venue.\cite{Bornmann2008,Wang2011} \tablename~\ref{tab:topcited} lists the ten most cited SESoS papers as recorded by Google Scholar regarding \added[id=EC]{total, organic, and average} number of citations. \added[id=EC]{The number of organic citations to a paper is computed by excluding author self-citations, i.e., citations to it within other papers for which they have at least one author in common.\cite{Aksnes2003}} In turn, the average number of citations is defined by the ratio between the absolute number of citations for a paper and the number of years since it has been published.\cite{Garousi2016} This treatment is important for a fair impact assessment since papers published longer ago are likely to obtain more citations.

\begin{center}
\begin{table*}
\caption{Top-ten cited SESoS papers as retrieved from Google Scholar  \added[id=EC]{(March 2024)}.\label{tab:topcited}}
\begin{tabular*}{\textwidth}{@{}lp{12cm}c>{\color{blue}}c>{\color{blue}}c@{}}
\multicolumn{5}{@{}l}{Ordered by total number of citations}\\[0.1cm]
\toprule
ID & Title & Year & Citations & Organic citations\\
\midrule
P8 & The state of the art and future perspectives in systems of systems software architectures & 2013 & 90 & 86\\
P6 & Towards an IoT ecosystem & 2013 & 66 & 66\\
P19 & On the characterization of missions of systems-of-systems & 2014 & 52 & 47\\
P21 & Quality attributes of systems-of-systems: A systematic literature review & 2015 & 48 & 48\\
P4 & Challenges for SoS architecture description & 2013 & 48 & 47\\
P1 & On the challenges of self-adaptation in systems-of-systems & 2013 & 43 & 42\\
P18 & Investigating the model-driven development for systems-of-systems & 2014 & 42 & 38\\
P20 & Characterizing architecture description languages for software-intensive systems-of-systems & 2015 & 36 & 35\\
P30 & Approaches for integration in system of systems: A systematic review & 2016 & 33 & 33\\
P14 & Live visualization of large software landscapes for ensuring architecture conformance & 2014 & 30 & 25\\
\bottomrule
\multicolumn{5}{c}{}
\end{tabular*}
\begin{tabular*}{\textwidth}{@{}lp{12.9cm}c>{\color{blue}}c@{}}
\multicolumn{4}{@{}l}{Ordered by average number of citations}\\[0.1cm]
\toprule
ID & Title & Year & Average citation count\\
\midrule
P55 & Integration challenges for digital twin systems-of-systems & 2022 & 13.00\\
P8 & The state of the art and future perspectives in systems of systems software architectures & 2013 & 8.18\\
P6 & Towards an IoT ecosystem & 2013 & 6.00\\
P21 & Quality attributes of systems-of-systems: A systematic literature review & 2015 & 5.33\\
P19 & On the characterization of missions of systems-of-systems & 2014 & 5.20\\
P46 & The status quo of systems-of-information systems & 2019 & 5.00\\
P38 & A meta-model for representing system-of-systems ontologies & 2018 & 4.67\\
P4 & Challenges for SoS architecture description & 2013 & 4.36\\
P18 & Investigating the Model-Driven Development for Systems-of-Systems & 2014 & 4.20\\
P35 & On the interplay of Business Process Modeling and missions in systems-of-information systems & 2017 & 4.14\\
\bottomrule
\end{tabular*}
\end{table*}
\end{center}

Considering the total number of citations, the top-cited SESoS papers concern issues related to SoS software architectures. Paper P8 reports a comprehensive SLR addressing the design, representation, evaluation, and evolution of these architectures, which are essential for contributing to the success of software-intensive SoS, besides identifying relevant research gaps in this context. Paper P4 and P20 also concern SoS software architectures with a specific focus on their architecture description and languages for this purpose. The ISO/IEC/IEEE 42010 International Standard\cite{ISO42010} acknowledges the critical role of architecture descriptions in SoS development, which can be used for improving communication, analysis, and evolution of these systems.

As mentioned in Section~\ref{subsec:rq2}, the community has demonstrated interest in quality issues for SoS. The SLR reported in paper P21 identified relevant quality attributes for SoS and limitations of existing quality models, confirming that addressing extra-functional properties in SoS still needs research efforts. The prevalence of interoperability observed in paper P21 is also object of study for paper 30, which concerns the integration of heterogeneous constituent systems into an SoS.

Another topic of interest refers to missions in SoS, which is the focus of paper P19. This paper presents a study about how the missions of SoS can be defined, specified, represented, and implemented. Missions could be the starting point to adopt when designing an SoS so that mission models can be used to drive the choice of constituent systems that shall be integrated into an SoS towards fulfilling its goals. In SESoS 2015, paper P23 (a step further in that research) pioneered the introduction of a language and tool to specify SoS missions. Even though several approaches and tools had been proposed for SoS modeling, they did handle mission modeling. On the other hand, approaches to handling missions in traditional software systems do not cope with the characteristics of SoS. The relevant role of missions has made this topic still an object of active research nowadays.\cite{Axelsson2022,Martin2023,Ferreira2023}

Despite its recentness, paper P55 achieved a significant number of citations in less than two years. This paper elicits and discusses some challenges related to the construction of digital twin SoS. One of the several existing definitions, a digital twin is a virtual representation of a physical object, process, or even a real-world system that is continuously synchronized throughout its life cycle.\cite{Eramo2022} Data managed by a digital twin about its entity can be analyzed, aggregated, and extended to support human-oriented or intelligent informed decisions, generate new information, and provide value-added services on these data to visualize, monitor, control, or optimize the observed entity. In this perspective, paper P55 presents a view of composing and integrating digital twins of smaller-scale systems to become constituents of an SoS. Digital twins have attracted growing research interest from academia and industry, but they have not yet been sufficiently investigated from a Software Engineering point of view.\cite{Rivera2020,Dalibor2022} The integration of heterogeneous digital twins to form an SoS exacerbates the existing challenges due to the embodiment of the distinguishing characteristics of SoS.

\begin{figure*}
\centerline{\includegraphics[scale=0.45]{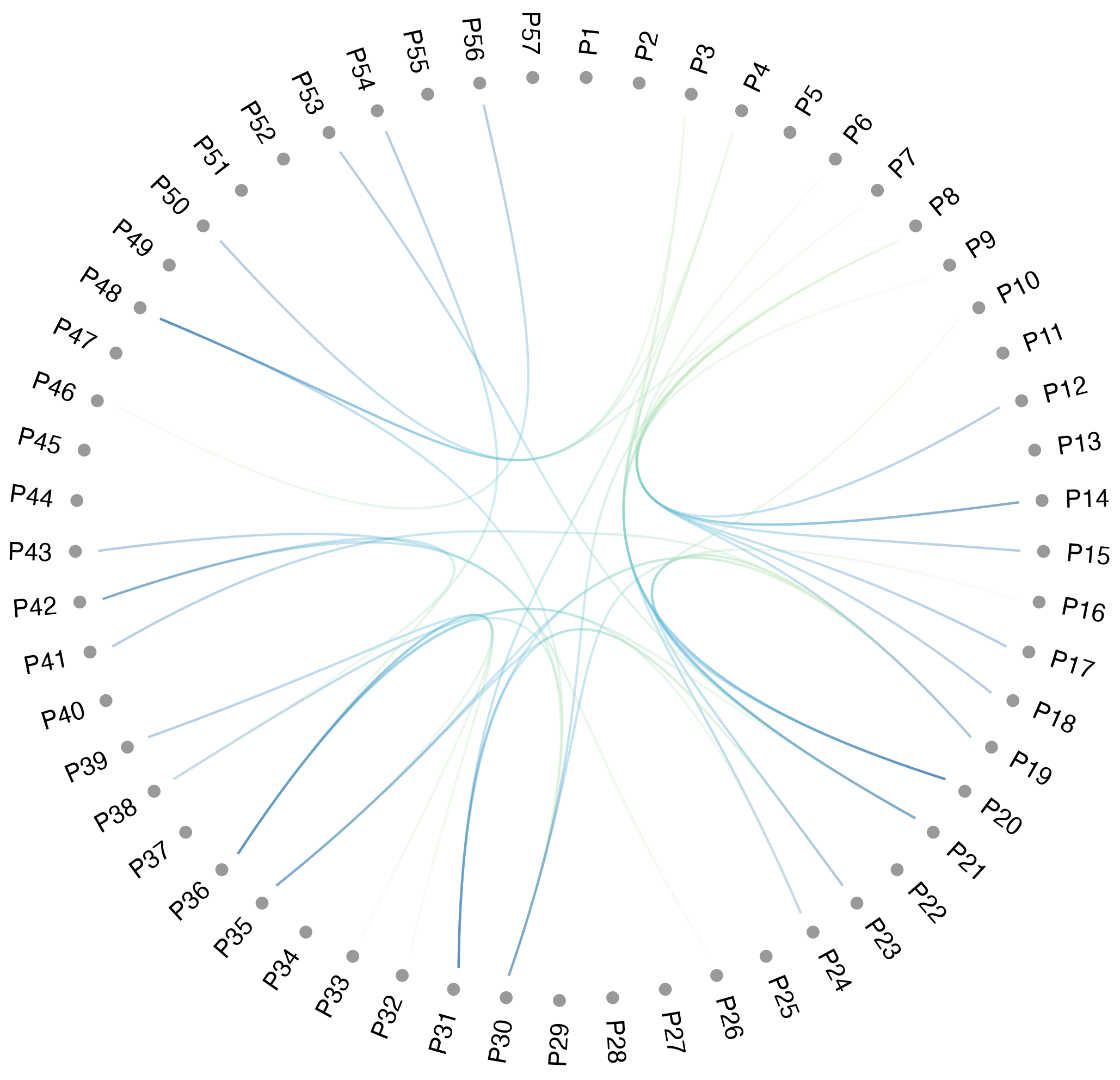}}
\caption{\blue{Hierarchical edge bundle representing citations between SESoS papers from 2013 to 2023.\label{fig:in-citations_sesos}}}
\end{figure*}

\blue{Another relevant facet regarding the impact of SESoS papers is assessing their citations within the community, i.e., how many papers published in recent editions of the workshop cite those appearing in past editions. \figurename~\ref{fig:in-citations_sesos} shows a hierarchical edge bundle representing citations between SESoS papers from 2013 to 2023. Hierarchical edge bundling is a tree-based visualization technique showing hierarchically organized nodes and connections (adjacency relations) between them. This technique reduces the visual clutter usually observed in complex networks, thus making those relations more discernible.\cite{Holten2006} In \figurename~\ref{fig:in-citations_sesos}, nodes (small circles) are papers, and links (colored lines\footnote{\added[id=EC]{In general, visualization techniques use colors to add information in a visual representation. The reader is invited to view a colored version of this article.}}) are citations between them. The links are colored to represent the direction of the citation: the citing paper is at the blue end of the link, and the cited paper is at the green end.}

\blue{From the hierarchical edge bundle shown in \figurename~\ref{fig:in-citations_sesos}, it is possible to notice that 20 out of 55 papers published in SESoS from 2013 to 2022\footnote{\added[id=EC]{The count considered the range 2013--2022 since there are no citations for papers published in 2023 within SESoS.}} ($\approx$ 36.4\%) were cited by other papers internally within the workshop's community. This means that a significant volume of research reported in SESoS has not been grounded on previous work in the workshop. On the other hand, the small proportion of citations within SESoS to the sum of the number of citations to the papers (39/1,068) indicates that the research reported at the workshop has been able to have a broader reach. It is also worth noticing that, from the 35 papers not cited within SESoS, 34.3\% of them ($n = 12$) are short/position papers. Therefore, a possible explanation for the significant number of non-cited SESoS papers is that regular papers report more solid research contributions; hence, they are more likely to be cited.}

\added[id=EC]{The most cited papers within the SESoS workshop are P8 (seven citations), P19 (five citations), and P4 (four citations). The discussions raised by paper P8, which is the most cited study in terms of total number of citations, have supported other subsequent studies related to SoS software architectures, such as those related to the representation (papers P12 and P20), construction (papers P30 and P50), and quality issues (paper P21) of those architectures. Paper P4, which is among the top-ten cited SESoS papers, sheds light on relevant issues to consider when describing SoS software architectures. In turn, paper P19 has played a relevant role in the community by reporting a study on missions of SoS, which are crucial for engineering these systems.}

\begin{mdframed}[innerleftmargin=0.25cm,innerrightmargin=0.25cm,innertopmargin=0.25cm,innerbottommargin=0.25cm,skipabove=0.5cm,linecolor=blue]
\added[id=EC]{Summary of findings:}
\begin{itemize}[leftmargin=*,topsep=-4pt]
	\item \added[id=EC]{The most cited SESoS papers concern SoS software architectures.}
	\item \added[id=EC]{The research community is interested in quality issues for SoS, especially interoperability.}
	\item \added[id=EC]{Addressing missions in SoS, including their specification and realization, represents a relevant research topic.}
	\item \added[id=EC]{Digital twin-based SoS is a topic that has drawn significant attention from the community in the last few years.}
	\item \added[id=EC]{A significant amount of the research reported at SESoS has not been grounded on previous work at the workshop, even though it has reached a broader community.}
\end{itemize}
\end{mdframed}

\subsection{\blue{Threats to validity}}
\label{subsec:threats-to-validity}
\added[id=EC]{This section discusses the main threats to the validity of the study reported in this article and the strategies to mitigate them. These threats are analyzed under the lens of the classification schema proposed by Ampatzoglou et al.\cite{Ampatzoglou2019} for reporting threats to validity in Software Engineering secondary studies and possible mitigation actions. Such a classification schema provides three categories, namely \textit{research validity}, \textit{study selection validity}, and \textit{data validity}.}

\added[id=EC]{\textbf{Research validity.} Threats in this category span the entire study as they are primarily related to how rigorous its design and conduction were, and issues concerning the study's generalizability. The scoping review reported in this article followed existing guidelines for conducting this kind of secondary study,\cite{Arksey2005,Peters2020} besides the fact that the researchers involved in this study are highly experienced in conducting secondary studies in Software Engineering and SoS. Furthermore, aiming at minimizing bias, the first stage of the study consisted of establishing a precise protocol with goals, research questions, eligibility criteria, and procedures for collecting, extracting data, and analyzing SESoS papers (see \figurename~\ref{fig:process}). Rigorously following that protocol increased the reliability and representativeness of the results in mapping the research reported in the SESoS workshop between 2013 and 2023.}

\added[id=EC]{The purposeful choice for the scoping review method, instead of an SLR or SMS, was to answer research questions regarding the nature and diversity of the knowledge (in this case, regarding Software Engineering for SoS) based on the analysis of SESoS papers. As mentioned in Section~\ref{sec:methodology}, this work followed Campbell et al.\cite{Campbell2023} in distinguishing a scoping review from SLRs and SMS, even though the similarities between these secondary studies often give room for misunderstanding. It is also worthwhile mentioning that scoping reviews have inherent limitations as they do not assess the rigor or quality of the studies they include.\cite{Arksey2005} A scoping review shall include all relevant literature regardless of a critique of the quality of those studies since it intends to present an overview of the research activity related to a topic of interest.\cite{Pham2014}}

\added[id=EC]{Even though SESoS is currently the main venue concerning Software Engineering for SoS, it is not the only one, and it is possible to find studies related to this topic elsewhere, e.g., in IEEE SoSE. Of course, this limits the generalization of the results because they cannot be regarded as mapping the entire landscape of Software Engineering for SoS, but they can be seen as a representative sample of it. Extrapolating the scoping review out of SESoS is not within the scope of this work, so future work can expand the study to other sources to complement or contrast the results currently reported here.}

\added[id=EC]{\textbf{Study selection validity.} Threats in this category are related to the validity of the study search and inclusion. A limitation often reported in several secondary studies is the possibility that the review may have missed some relevant studies for many reasons (e.g., inadequate selection of sources, deficiencies in constructing search string, study selection bias, etc.), compromising its completeness and coverage of the literature. However, this limitation does not apply to this study since the stage regarding identifying relevant studies (see \figurename~\ref{fig:process}) was delimited to papers published in the SESoS workshop. The probability of missing studies is negligible because no constraints were imposed to retrieve those studies, and the ACM Digital Library and IEEE Xplore are expected to have correctly indexed all the papers in the workshop proceedings. Study selection bias could also be a threat to the study selection validity, but the exclusion criterion adopted for this study is simple in the sense that it concerns excluding only papers unrelated to SoS.}

\added[id=EC]{The snowballing technique,\cite{Wohlin2014} which uses the reference list of a study (backward snowballing) or citing references to it (forward snowballing) to identify additional studies, is often used to increase literature coverage in secondary studies, thus contributing to minimizing threats to study selection validity. This study did not use snowballing because the analysis corpus was delimited to SESoS. Recall that the goal of the scoping review reported in this article was to categorize research in Software Engineering for SoS specifically presented in SESoS, so exploring other venues than this workshop did not specifically pertain to the goal of this work.}

\added[id=EC]{\textbf{Data validity.} Threats in this category are primarily related to data extraction and analysis processes. Data extraction bias refers to any issue that can be introduced while collecting data, namely bias, inaccuracies, subjectivity in quality assessment, unverified data extraction, and study misclassification. Not all data extracted from SESoS papers to answer the research questions were obvious, and some data had to be inferred, thus opening the possibility for the researchers to introduce bias even unintentionally. To minimize imprecision and bias in data collection, analysis, and synthesis, obtained data and consequent results were reviewed among the researchers involved in this study, who have been working with Software Engineering and SoS for several years, and possible disagreements have been solved to reach consensus. Furthermore, the classification of SESoS papers into the categories defined in the ACM CCS, the topics of interest of the workshop, and the research types was preceded by content analysis relying on open coding so that the codes identified from the goals and contributions of each paper allowed for minimizing misclassification. It is also worth highlighting that the classification schema provided by ACM CCS is recognized worldwide, and hundreds of studies in the Software Engineering literature have been using the taxonomy of research types proposed by Wieringa et al.\cite{Wieringa2006}}

\added[id=EC]{Scientometrics-based data analysis is not also free from shortcomings. Grinäv\cite{Grinav2020} argues that the leading scientometric indicators, such as the number of citations and derived indexes, might not provide an objective picture of the actual contribution or impact of a study, researcher, or institution. The scientometric analysis carried out in the final stage of the scoping review (see \figurename~\ref{fig:process}) considered citation analysis to assess the significance, utility, attention, visibility, or short-term impact of SESoS papers. The literature widely uses citation count as a metric for measuring impact mainly due to its straightforwardness, but, at the same time, it acknowledges that it can be influenced by many confounding factors and a diversity of motivations for citing a study.\cite{Bornmann2008,Wang2011,Molleri2018} To provide a more fair impact assessment, the citation analysis considered not only the total number of citations reported by Google Scholar but also counted the average number of citations for the papers' age and the number of citations excluding self-citations (see Section~\ref{subsec:threats-to-validity}). The exclusion of self-citations is also imperfect since there are genuine reasons for an author to cite his/her own work, such as the incremental nature of research and the dependency or relationship of the present work on previous ones. In addition, it is worthwhile mentioning that citation counts likely increase over the years, but such an increase is not linear.\cite{Molleri2018}}

\added[id=EC]{Using Google Scholar as the source for obtaining the number of citations of SESoS papers is also debatable since it does not provide selection criteria based on types of publications (unpublished and non-scholarly works are considered) and may present inaccuracies.\cite{Grinav2020} On the other hand, the Google Scholar engine is free and uses a broad range of data sources. As Google Scholar can retrieve grey literature, which is also an acknowledged source of evidence in Software Engineering,\cite{Garousi2020} sometimes the number of publications listed there is greater than that found in well-known electronic databases such as Scopus and Web of Science, thereby increasing the total number of citations. Nevertheless, using Web of Science, for example, would not exempt citation analysis from problems, such as closed access, underestimation of citation impact, and constraints on considering only the sources it indexes.\cite{Harzing2008} The results about the impact of SESoS papers based on citation analysis shall hence be not taken as an absolute truth but instead critically scrutinized.}

\section{Some current and future research directions in Software Engineering for Systems-of-Systems}
\label{sec:directions}
\begin{figure*}
\centerline{\includegraphics[width=\textwidth]{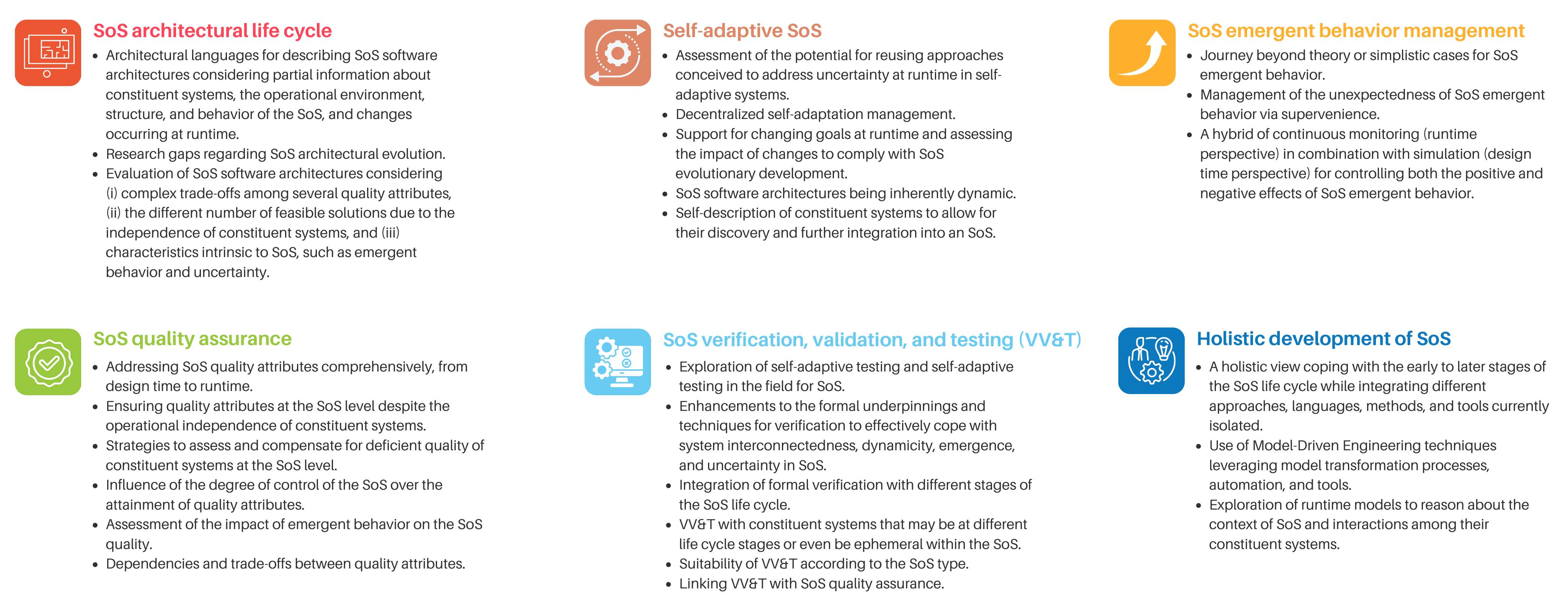}}
\caption{\blue{Highlights of some current and future research directions in Software Engineering for SoS.}\label{fig:research-directions}}
\end{figure*}

\added[id=EC]{The inherently complex nature and unique characteristics of SoS, such as the independence of their constituent systems and (unexpected) emergent behavior stemming from the interactions among them, make designing SoS a task far from trivial. This section delves into a non-exhaustive set of relevant research directions in Software Engineering for SoS. They primarily derive from the research gaps raised by the 16 papers classified under the \textit{Challenges \& Directions} category of topics of interest for SESoS, including nine secondary studies (see Section~\ref{subsec:rq2}). \figurename~\ref{fig:research-directions} summarizes these directions.}

\subsection{\blue{SoS architectural life cycle}}
\added[id=EC]{The major research challenges raised by software-intensive SoS are fundamentally architectural.\cite{Cadavid2020} The significant relevance of software architectures for sucessfully designing and developing SoS is evidenced by the number of papers addressing this topic in SESoS (see Section~\ref{subsec:rq2}), the many other studies in the literature (see Section~\ref{subsec:rq7}), and the existence of an international standard embracing SoS software architectures.\cite{ISO42010} The SoS architectural life cycle mapped into an overarching process and supported by tools should consider the inherent characteristics of SoS in terms of partial information about constituent systems, emergent behavior, and dynamicity. These issues can give rise to different research threads related to SoS software architectures, specifically their representation, evaluation, and evolution.}

\added[id=EC]{Work on the representation of SoS software architectures has been concerned with assessing the adequacy of existing architectural languages and the required viewpoints and level of formalism, as identified by the SLR reported in paper P8. Paper P4 argues that there is a need for proper architectural languages to cope with the characteristics of SoS, including the definition of architectural elements and their interactions and support for dynamicity and quality issues. In turn, paper P20 identifies some essential features that architectural languages should provide to describe SoS software architectures. Guessi et al.\cite{Guessi2015} carried out an SLR on the description of SoS software architectures and identified a variety of architectural languages for this purpose. Although all these works have demonstrated research advances in the representation of SoS software architectures, there are still gaps to be addressed, particularly regarding the independence of constituent systems and the evolutionary development of SoS. SoS software architectures shall consider partial, mutable information about constituent systems and the operational environment, structure, and behavior of the SoS. Adopting traditional approaches and languages for the architectural description of SoS cannot properly consider these features. A way of tackling these issues when representing SoS software architectures could be using an intentional, abstract approach in which constituent systems to form an SoS are declared based on their types and capabilities to be further concretized at runtime,\cite{Oquendo2016b} or in terms of constraints governing the synthesis of those architectures.\cite{Guessi2016,Guessi2020}}

\added[id=EC]{By 2013, paper P8 concluded that the evaluation and evolution of SoS software architectures deserved attention and further research effort. The gap regarding the architectural evolution of SoS remains after ten years. Surprisingly, the literature on this topic is still scarce despite evolutionary development being one of the fundamental characteristics of SoS, and their software architectures must cope with it. On the other hand, there have been significant advances in the architectural evaluation of SoS in the last decade. Santos et al.\cite{Santos2023b} report the results of a recent SLR on the state-of-the-art regarding how SoS architectures have been evaluated. Their findings show that (i) most studies proposed or used evaluation methods relying on mathematical modeling and simulation in contrast with well-known scenario-based methods used in the architectural evaluation of traditional software-intensive systems, and (ii) evaluation methods for SoS software architectures need consolidation and flexibility. Nonetheless, the characteristics of SoS pose many relevant challenges when it comes to architectural evaluation. Architectural evaluation in SoS must consider several quality attributes, thereby making the analysis of dependencies and trade-offs more complex since the operational independence of constituent systems directly impacts if and how an SoS achieves those quality attributes. The independence of constituent systems, which may join or leave an SoS at any time, also affects the number of feasible SoS architectural solution alternatives to evaluate and how they can be evaluated under uncertainties. Furthermore, emergent behavior, which is also vital in any SoS, can lead to a complete breakthrough in architectural evaluation in this context due to its features and implications.}

\subsection{\blue{Self-adaptive SoS}}
\added[id=EC]{Contemporary systems, including SoS, are expected to maintain availability even in face of uncertainties and operation under dynamic and changing environments. SoS are not exempt from these features. In SoS, dynamicity mainly stems from the fact that constituent systems can be integrated, operated, and reconfigured at runtime, often in an unplanned way and out of the control of the SoS. The operational and managerial independence of constituent systems of SoS impose challenges related to this dynamicity since such systems are independent entities that can evolve independently without the control of the SoS, thus impacting their structure and behavior. Therefore, supporting this dynamicity is essential, especially in the case of certain critical systems, such as those in air traffic control, energy, disaster management, environmental monitoring, and health care. SoS in these scenarios must maintain high availability, thereby requiring dynamic reconfigurations.}

\added[id=EC]{The dynamicity observed in SoS makes these systems to be self-adaptive by nature. A self-adaptive system is a system able to modify its structure or behavior in response to its perception of the environment, the system itself, and its goals.\cite{DeLemos2013} While self-adaptation seems well-established in the field, many solutions for engineering self-adaptive systems are often complex. The most recent research advances related to self-adaptive systems have primarily focused on uncertainties arising from the lack of complete knowledge of the system and its executing conditions before deployment, as well as how to control them at runtime.\cite{Weyns2019} A significant body of knowledge about self-adaptive systems has been accumulated over the last three decades,\cite{Krupitzer2015,Wong2022} with roadmaps outlining research challenges structured along different aspects of engineering these systems.\cite{DeLemos2013,Cheng2009,DeLemos2017} Nevertheless, little is known to what extent this existing knowledge can be applied to SoS.}

\added[id=EC]{The Monitor--Analyze--Plan--Execute over a shared Knowledge (MAPE-K) feedback loop\cite{Kephart2003} has been the de facto reference control model in engineering self-adaptive systems.\cite{Arcaini2015} However, it assumes a centralized perspective that may not fit SoS, which are inherently decentralized. Weyns\cite{Weyns2019,Weyns2021} points out that adaptation in decentralized settings represents a current research challenge for self-adaptive systems, which also applies to SoS. Much of the research to date has considered self-adaptive systems maintaining stated goals, even in uncertain and changing environments, but such goal stability is undoubtedly not the case for SoS. Therefore, SoS with self-adaptive features should support changing goals at runtime to comply with their inherent evolutionary development and assess the impact of those changes on constituent systems and the overarching SoS. Paper P9 highlights that the inherent dynamicity of SoS drives the need for runtime monitoring and the satisfaction of requirements, but the specific SoS type must also be considered. A possible approach for these issues might find inspiration in requirements reflection, which considers requirements as runtime entities as means of enabling systems to reason about, understand, explain, and modify requirements at runtime.\cite{Sawyer2010,Bencomo2010}}

\added[id=EC]{Finally, a particularity of SoS software architectures is their inherent dynamicity. They must adapt or evolve at runtime due to the operational and managerial independence of constituent systems and the evolutionary development and emergent behavior in the SoS. Although dynamicity is inherent to SoS, the literature on dynamic SoS software architectures is still limited,\cite{Heinrich2023,Manzano2020} thus highlighting the need to advance research in this context. For SoS, it is crucial to understand why, how, when, and where such changes occur and what characteristics and decisions should be made to avoid degradation of the architecture, functionalities, and quality attributes of these systems. Furthermore, paper P56 pinpoints that constituent systems should self-describe themselves to allow for their discovery and seamless integration into an SoS. These features are essential because the constituent systems comprising the SoS may only be partially known beforehand, and alternative systems may also contribute to fulfilling its missions. As a result, reconfigurations may be necessary at runtime to accommodate these dynamics and ensure the continued effectiveness of the SoS.}

\subsection{\blue{SoS emergent behavior management}}
\added[id=EC]{SoS exhibit emergent behavior resulting from the local interaction among their independent constituent systems and rely on it to offer unique capabilities. This characteristic is distinctive from what is observed in traditional systems, whose behavior can be understood as the sum of the behavior of their constituent elements (components and sub-systems). However, this reductionism fails in SoS because constituent systems may behave in ways that cannot be predicted exclusively from analyzing their individual behavior. The interactions among constituent systems may also be ephemeral due to their operational independence, which may impact their availability in an SoS. Consequently, the emergent behavior of an SoS is unexpected.\cite{Oquendo2018a} The ultimate aim of an SoS is to achieve its missions while attempting to ensure that the interactions among its constituent systems do not yield an undesirable behavior as a side-effect. On the other hand, SoS in safety-critical domains do not tolerate partially known or unknown emergent effects that can cause a deviation of the actual behavior from the intended behavior at the cost of catastrophic consequences.\cite{Kopetz2016}}

\added[id=EC]{SoS emergent behavior management remains a significant challenge from both Systems Engineering and Software Engineering perspectives. While being vital for any SoS, emergent behavior is perhaps the least understood aspect. Much of the existing research focuses on characterizing emergent behavior in SoS and its implications, but these efforts remain mostly theoretical.\cite{Inocencio2019} Oquendo\cite{Oquendo2018a} offers a valuable contribution by discussing the concept of emergence in the context of SoS engineering. He explains that SoS exhibit systemic emergence in which the emergent behavior of an SoS arises from the supported interactions among its constituent systems and can be deducible, even if unexpected. He further argues that supervenience is a well-founded approach to cope with the unexpectedness of emergent behavior in SoS. From the supervenience point of view, the emergent behavior of the SoS lies at a global macro-scale level, whereas the behavior of its constituent systems is at an individual micro-scale level. An upward causation from the micro-scale level to the macro-scale level makes changes in interacting constituent systems' behavior to imply changes in the emergent behavior of the SoS. Therefore, exploring the application of supervenience principles in the design and operation of software-intensive SoS presents a promising avenue for future research, particularly from the Software Engineering perspective.}

\added[id=EC]{Paper P9 acknowledges the challenges of predicting SoS behavior and that some issues related to the interactions among constituent systems may only manifest in the SoS after a prolonged operation. The authors also advocate for continuous monitoring of the SoS operation to identify potentially critical performance situations or events, including deviations from the expected or desirable behavior. To achieve this goal, the authors propose instrumenting an SoS with probes to capture relevant system events and data at both the SoS and constituent systems levels, including their interactions. This idea aligns with paper P56, which emphasizes controlling (visualizing, mastering, validating) the emergent behavior of an SoS to mitigate undesirable effects. While a monitoring approach appears feasible as it addresses a runtime perspective, further research must consider the SoS type and decisions about what, when, where, and how to monitor and which entity would be responsible for this task. Conversely, design-time approaches such as modeling and simulation have been extensively studied in SoS engineering, yielding substantial insights and results.\cite{Nielsen2015,GracianoNeto2018,Zeigler2019} A possible research direction could be exploring a hybrid approach that integrates design-time and runtime aspects. By leveraging the strengths of each approach, such hybrid models can enhance the understanding of SoS behavior and improve the control over both positive and negative effects of emergence.}

\added[id=EC]{Kounev et al.\cite{Kounev2017} define a self-aware system as one that (i) comprises models that continuously capture knowledge about the system itself and its environment in terms of its structure, state, possible actions, and behavior at runtime and (ii) leverages these models to predict, analyze, and plan actions based on that knowledge. Self-awareness has been employed to enable self-adaptation in software-intensive systems exhibiting uncertain and dynamic behaviors,\cite{Elhabbash2019} but a critical limitation is that a self-adaptive system knows only about itself rather than its interactions with other systems. On the other hand, SoS must have collective (self-)awareness. This means that an SoS should have knowledge of the interactions among its constituent systems as they give rise to the emergence of a behavior essential for achieving its missions. A research direction could be bringing self-awareness to effectively manage emergent behavior in SoS, a perspective that still needs to be further explored in the literature.\cite{Motus2012,Cavalcante2015}}

\subsection{\blue{SoS quality assurance}}
\added[id=EC]{SoS development requires quality at the forefront due to the critical nature of these systems in many scenarios. As evidenced in Section~\ref{subsec:rq2}, the research community has recognized the importance of SoS quality, especially the interoperability among constituent systems fostering seamless interaction. Papers P48 and P50 in SESoS aimed at promoting the integration of constituent systems through their software architectures in the early stages of the SoS life cycle. However, it is noteworthy that most existing approaches primarily focus on design-time constituent system interoperability and overlook or even neglect the highly dynamic, evolving nature of SoS from a runtime point of view. This imperative extends to other pivotal quality attributes, e.g., security.\cite{ElHachem2016,Olivero2019,Olivero2023} A recent contribution in this direction is the ReViTA framework, introduced by Ferreira et al.,\cite{Ferreira2024} which explicitly focuses on ensuring dependability in SoS at runtime through architectural dynamic reconfigurations. Their authors also note that changing a given constituent system for another at runtime may degrade the SoS quality in terms of performance, security, etc., with the new architectural configuration potentially needing to sort quality attributes differently from the original one. Therefore, addressing quality attributes comprehensively throughout the SoS life cycle becomes paramount, spanning from design time to runtime considerations.}

\added[id=EC]{The inherent characteristics of SoS present significant challenges for quality assurance, giving room to several questions deserving further investigation. How can quality attributes (performance, safety, security, etc.) be reliably ensured at the SoS level despite the operational independence of constituent systems? How can the SoS keep a high-quality operational standard when a constituent system joins or leaves the SoS? How do constituent system deficiencies with respect to a given quality attribute impact SoS-level achievement of that attribute? If a constituent system fails to uphold a particular quality attribute, how can the SoS compensate to achieve it? How does the degree of control of the SoS over its constituent systems (based on the SoS type) influence the difficulty of attaining desired quality levels? How do interactions among constituent systems (and the consequent emergent behavior) affect the SoS quality? How can dependencies and trade-offs between quality attributes be effectively analyzed, considering the inherent complexity of SoS? What are the implications of trading off quality attributes across the SoS? These questions underscore the multifaceted nature of quality assurance in SoS and highlight the need for comprehensive investigation and innovative solutions.}

\subsection{\blue{SoS verification, validation, and testing}}
\added[id=EC]{The analysis of papers that appeared in SESoS from 2013 to 2023 revealed a significant focus on SoS VV\&T (see Sections~\ref{subsec:rq2} and~\ref{subsec:rq3}). There are several reasons why these processes are crucial for SoS. First, VV\&T ensure the correct and seamless collaboration between the heterogeneous constituent systems toward achieving SoS missions. Second, they help mitigate undesirable behavior, thus enhancing the SoS reliability and expected operation. At last, they foster the continued operation and quality of the SoS amidst changes in the constituent systems and the environment, besides making it possible to assess the impact of such changes.}

\added[id=EC]{While the role of VV\&T in SoS is widely acknowledged, the distinguishing characteristics of these systems raise significant challenges demanding further research efforts.\cite{OliveiraNeves2018} One of the crucial questions arising in this context is: how can a system undergo VV\&T when its constituent parts are only partially known or entirely unknown? Moreover, SoS as open-world systems lacking complete information about their operational environment complicate the definition of an a priori test environment. Even though simulation has often been used in the context of SoS,\cite{GracianoNeto2017} its effectiveness may be limited by the complexity and intricacies of emergent behavior.}

\added[id=EC]{The dynamic nature of SoS, characterized by the absence of a steady state, raises further questions regarding how to conduct VV\&T effectively. Self-adaptive system testing has recently provided promising directions to tackle dynamic and uncertain scenarios,\cite{Fredericks2014}, but its applicability to SoS requires further investigation. Another quite recent but promising concept for SoS is self-adaptive testing in the field,\cite{Silva2024} which dynamically adjusts testing based on system evolution, environmental changes, user behavior, new functionalities, or emerging failures. This approach appears particularly relevant for SoS due to their interconnected, evolving nature and operation in uncertain environments with limited control. Exploring this adaptive approach could offer valuable insights into addressing the unique testing challenges inherent in SoS.}

\added[id=EC]{Formal verification is imperative for mission-critical domains, such as defense, healthcare, and transportation, where unintended events in SoS may lead to severe consequences. For this purpose, the literature has investigated traditional model checking and the more computationally efficient statistical model checking (SMC). SMC is a probabilistic, simulation-based technique that intends to verify, at a given confidence level, if a certain property is satisfied during the system's execution, balancing accuracy and computational effort.\cite{Legay2010} Unlike conventional model checking relying on the analysis of the internal logic of the target system, SMC analyzes the traces representing the system's execution states against the properties to be verified. These features make SMC applicable even when the system's internals are inaccessible or unknown and its behavior is not fully understood or controllable,\cite{Agha2018} as it is the case for SoS and their constituent systems since a sufficient condition for using SMC is that the system under verification is executable. Although SMC has been proven effective in SoS,\cite{Legay2015,Seo2016} the formal underpinnings typically used to specify the properties to be verified, such as linear temporal logic and derivatives, do not consider dynamicity. These limitations hinder the application of those formalisms to SoS due to the emergence of interactions among constituent systems and the fact that a given constituent system may be within the SoS in a particular instant of time and no longer be there at another. Some work in the literature has introduced a new logic to address dynamicity.\cite{Cavalcante2016,Quilbeuf2016} Nevertheless, current formal methods still need enhancements to cope with system interconnectedness, dynamicity, emergence, and uncertainty,\cite{Michael2020} which are inherent to SoS. Another relevant research thread is not only performing a standalone formal verification but also integrating it with different stages of the SoS life cycle, e.g., leveraging mission models and software architecture descriptions.\cite{Oquendo2018b,Mohsin2020,Silva2020}}

\added[id=EC]{The inherent characteristics of SoS introduce other concerns regarding the VV\&T of these systems. The managerial independence of constituent systems and the evolutionary development of SoS raise questions about verifying, validating, and testing constituent systems that may be at different life cycle stages or even be ephemeral within the SoS. Moreover, it is necessary to consider how SoS types influence the approach(es) to VV\&T and how these processes can be effectively linked with the assurance of SoS quality attributes.}

\subsection{\blue{Holistic development of SoS}}
\added[id=EC]{Research on Software Engineering for SoS has not yet achieved a holistic view of coping with the early to later stages of the SoS life cycle while integrating different approaches, languages, methods, and tools that are currently isolated. This holistic view could start from mission models and support SoS design, development, and evolution while assuring the achievement of quality concerns (such as dependability, scalability, security, safety, and others) and addressing trade-offs among them. Dynamicity should also be considered since it is a challenging dimension for the SoS ever-changing context where constituent systems operate independently and can dynamically join and leave an SoS over time. Moreover, the scenario of designing, operating, and evolving SoS encompasses inherent uncertainty, complexity, interconnectedness, and emergence, which differ from what current Systems Engineering and Software Engineering methods and tools are used to address.}

\added[id=EC]{A promising research thread is to explore model-driven engineering (MDE) techniques to encompass transformation processes among the different models, automate the mapping process, and provide automated tools that hide the burden of dealing with the SoS intricacies. Paper P18 presented the state of the art regarding using model-driven techniques for SoS, advocating that this approach could fit these systems due to its success in Software Engineering in taming software complexity through high-abstraction models. The use of MDE techniques is an active topic in SoS engineering,\cite{Nielsen2015} which can be corroborated by the fact that almost a fifth of the analyzed SESoS papers have considered this approach (see Section~\ref{subsec:rq2}). Nevertheless, these studies have primarily concerned design-time modeling, which is only a piece of a broader context of using models in the life cycle of SoS and their constituent systems. Models reasoning about the context of SoS and interactions among their constituent systems at runtime are still under-explored for the case of SoS.\cite{Bencomo2019}}

\section{Conclusion}
\label{sec:conclusion}
This article presented a comprehensive landscape of the research on Software Engineering for SoS specifically reported in the annual editions of the SESoS workshop, a relevant scientific forum that has fostered the SoS community since 2013. The work reported in this article followed an evidence-based approach relying on scoping review and scientometric analysis methods to bring a historical perspective of SESoS. The study defined seven research questions and carefully analyzed 57 papers about SoS among the 94 published in SESoS proceedings, covering all workshop editions (2013 to 2023). The obtained results include the temporal distribution of the papers along the SESoS editions, the elucidation of top contributing countries to SESoS and co-authorship networks, the prevailing topics, research methodologies employed, application domains tackled, SoS types, and an analysis of the research impact of SESoS papers. All this information brings a consistent, detailed panorama of Software Engineering for SoS through the lens of the SESoS workshop.

The main contributions of this article are twofold. First, it offers a comprehensive landscape of what has been done regarding Software Engineering for SoS by analyzing the research reported in the SESoS workshop. This analysis delves into various aspects, including the temporal distribution and geographic location of research, prominent topics, research methodologies employed, application domains tackled, addressed SoS types, and research impact. Second, by identifying research gaps within this landscape, this article highlights different topics requiring further investigation in Software Engineering for SoS.

The combination of the distinctive characteristics of SoS amplifies the inherent complexity of designing, analyzing, developing, maintaining, managing, and evolving these systems. These characteristics bring several challenges from the Software Engineering point of view as the mainstream body of knowledge has been primarily focusing on single systems, not on SoS.\cite{Oquendo2016a} Therefore, this article also discussed  research directions related to Software Engineering for SoS, which are critical to advancing the ability to design, implement, and maintain high-quality, reliable SoS. These directions concern the design of SoS able to accommodate changes, ensure high quality standards, and respond to emergent behavior across the life cycle of these systems.

\added[id=EC]{Finally, the retrospective analysis presented in this article also intended to offer an overview of the evolution of SESoS over time and pinpoint potential enhancements for upcoming workshop editions. Despite establishing itself as a prominent venue for presenting and discussing Software Engineering research advances for constructing software-intensive SoS, SESoS is still predominantly an academic forum. The findings presented in this article underscored the need for more involvement of the industry in the research showcased at SESoS. The study also revealed the fragmentation within the SESoS research community and the need for further international collaborations among research groups working on similar or identical topics. Therefore, the SESoS workshop should explore ways to foster collaboration, aligning with its mission of leveraging the community's research endeavors.}

\bmsection*{Author contributions}
\textbf{E. Cavalcante:} Conceptualization, Methodology, Formal analysis, Investigation, Writing - Original Draft, Visualization. \textbf{T. Batista:} Writing - Review \& Editing. \textbf{F. Oquendo:} Writing - Review \& Editing.

\bmsection*{\blue{Data availability}}
\label{ex:data-availability}
\added[id=EC]{The electronic spreadsheet recording all the data extracted from SESoS papers is publicly available online at \url{https://bit.ly/sesospapers-datasheet}.}

\bmsection*{Financial disclosure}
None reported.

\bmsection*{Conflict of interest}
The authors declare no potential conflict of interest.

\bibliography{references}


\appendix

\bmsection{Papers on systems-of-systems in SESoS 2013-2023\label{app:papers}}
\vspace*{6pt}

\scrisize
\setlength{\baselineskip}{11pt}
\begin{enumerate}[label={[P\arabic*]},leftmargin=1cm,itemsep=4pt]
\item Weyns D, Andersson J. On the challenges of self-adaptation in systems of systems. In: First International Workshop on Software Engineering for Systems-of-Systems. ACM; 2013. \pdoi{10.1145/2489850.2489860}
\item Lytra I, Zdun U. Supporting architectural decision making for systems-of-systems design under uncertainty. In: First International Workshop on Software Engineering for Systems-of-Systems. ACM; 2013. \pdoi{10.1145/2489850.2489859}
\item Pérez J, Díaz J, Garbajosa J, Yagüe A, Gonzalez E, Lopez-Perea M. Large-scale smart grids as system of systems. In: First International Workshop on Software Engineering for Systems-of-Systems. ACM; 2013. \pdoi{10.1145/2489850.2489858}
\item Batista T. Challenges for SoS architecture description. In: First International Workshop on Software Engineering for Systems-of-Systems. ACM; 2013. \pdoi{10.1145/2489850.2489857}
\item Romay MP, Cuesta CE, Fernández-Sanz L. On self-adaptation in systems-of-systems. In: First International Workshop on Software Engineering for Systems-of-Systems. ACM; 2013. \pdoi{10.1145/2489850.2489856}
\item Delicato FC, Pires PF, Batista T, Cavalcante E, Costa B, Barros T. Towards an IoT ecosystem. In: First International Workshop on Software Engineering for Systems-of-Systems. ACM; 2013. \pdoi{10.1145/2489850.2489855}
\item Papatheocharous E, Axelsson J, Andrersson J. Issues and challenges in ecosystems for federated embedded systems. In: First International Workshop on Software Engineering for Systems-of-Systems. ACM; 2013. \pdoi{10.1145/2489850.2489854}
\item Nakagawa EY, Gonçalves M, Guessi M, Oliveira LBR, Oquendo F. The state of the art and future perspectives in systems of systems software architectures. In: First International Workshop on Software Engineering for Systems-of-Systems. ACM; 2013. \pdoi{10.1145/2489850.2489853}
\item Vierhauser M, Rabiser R, Grünbacher P, Danner C, Wallner S. Evolving systems of systems. In: First International Workshop on Software Engineering for Systems-of-Systems. ACM; 2013. \pdoi{10.1145/2489850.2489851}
\item Belloir N, Chiprianov V, Ahmad M, Munier M, Gallon L, Bruel JM. Using RELAX operators into an MDE security requirement elicitation process for systems of systems. In: 2014 European Conference on Software Architecture Workshops. ACM; 2014. \pdoi{10.1145/2642803.2642835}
\item Schneider JP, Teodorov C, Senn E, Champeau J. Towards a dynamic infrastructure for playing with systems of systems. In: 2014 European Conference on Software Architecture Workshops. ACM; 2014. \pdoi{10.1145/2642803.2642834}
\item Khlif I, Kacem MH, Kacem AH, Drira K. A multi-scale modelling perspective for SoS architectures. In: Proceedings of the 2014 European Conference on Software Architecture Workshops. ACM; 2014. \pdoi{10.1145/2642803.2642833}
\item Amorim SdS, de Almeida ES, McGregor JD, von Flach Chavez C. When ecosystems collide. In: 2014 European Conference on Software Architecture Workshops. ACM; 2014. \pdoi{10.1145/2642803.2642832}
\item Fittkau F, Stelzer P, Hasselbring W. Live visualization of large software landscapes for ensuring architecture conformance. In: 2014 European Conference on Software Architecture Workshops. ACM; 2014. \pdoi{10.1145/2642803.2642831}
\item Axelsson J, Kobetski A. Architectural concepts for federated embedded systems. In: 2014 European Conference on Software Architecture Workshops. ACM; 2014. \pdoi{10.1145/2642803.2647716}
\item Malakuti S. Detecting emergent interference in integration of multiple self-adaptive systems. In: Proceedings of the 2014 European Conference on Software Architecture Workshops. ACM; 2014. \pdoi{10.1145/2642803.2642826}
\item Maia P, Cavalcante E, Gomes P, Batista T, Delicato FC, Pires PF. On the development of systems-of-systems based on the Internet of Things. In: 2014 European Conference on Software Architecture Workshops. ACM; 2014. \pdoi{10.1145/2642803.2642828}
\item Graciano Neto VV, Guessi M, Oliveira LBR, Oquendo F, Nakagawa EY. Investigating the model-driven development for systems-of-systems. In: 2014 European Conference on Software Architecture Workshops. ACM; 2014. \pdoi{10.1145/2642803.2642825}
\item Silva E, Cavalcante E, Batista T. On the characterization of missions of systems-of-systems. In: 2014 European Conference on Software Architecture Workshops. ACM; 2014. \pdoi{10.1145/2642803.2642829}
\item Guessi M, Cavalcante E, Oliveira LBR. Characterizing architecture description languages for software-Intensive systems-of-systems. In: 2015 IEEE/ACM 3rd International Workshop on Software Engineering for Systems-of-Systems. IEEE; 2015. \pdoi{10.1109/sesos.2015.10}
\item Bianchi T, Santos DS, Felizardo KR. Quality attributes of systems-of-systems: A systematic literature review. In: 2015 IEEE/ACM 3rd International Workshop on Software Engineering for Systems-of-Systems. IEEE; 2015. \pdoi{10.1109/sesos.2015.12}
\item Petitdemange F, Borne I, Buisson J. Approach based patterns for system-of-systems reconfiguration. In: IEEE/ACM 3rd International Workshop on Software Engineering for Systems-of-Systems. IEEE; 2015. \pdoi{10.1109/sesos.2015.11}
\item Silva E, Batista T, Cavalcante E. A mission-oriented tool for system-of-systems modeling. In: IEEE/ACM 3rd International Workshop on Software Engineering for Systems-of-Systems. IEEE; 2015. \pdoi{10.1109/sesos.2015.13}
\item Perez J, Diaz J, Garbajosa J, Yague A, Gonzalez E, Lopez-Perea M. Towards a reference architecture for large-scale smart grids system of systems. In: IEEE/ACM 3rd International Workshop on Software Engineering for Systems-of-Systems. IEEE; 2015. \pdoi{10.1109/sesos.2015.9}
\item Nakagawa EY, Oquendo F, Avgeriou P, et al. Foreword: Towards reference architectures for systems-of-systems. In: IEEE/ACM 3rd International Workshop on Software Engineering for Systems-of-Systems. IEEE; 2015. \pdoi{10.1109/sesos.2015.8}
\item Seo D, Shin D, Baek YM, et al. Modeling and verification for different types of system of systems using PRISM. In: 4th International Workshop on Software Engineering for Systems-of-Systems. ACM; 2016. \pdoi{10.1145/2897829.2897833}
\item Salvaneschi P. Modeling of information systems as systems of systems through DSM. In: 4th International Workshop on Software Engineering for Systems-of-Systems. ACM; 2016. \pdoi{10.1145/2897829.2897832}
\item Park S, Park YB. ITE Arbitrator: A reference architecture framework for sustainable IT ecosystems. In: 4th International Workshop on Software Engineering for Systems-of-Systems. ACM; 2016. \pdoi{10.1145/2897829.2897834}
\item Johnson P, Lagerström R, Ekstedt M, Franke U. Modeling and analyzing systems-of-systems in the multi-attribute prediction language (MAPL). In: 4th International Workshop on Software Engineering for Systems-of-Systems. ACM; 2016. \pdoi{10.1145/2897829.2897830}
\item Vargas IG, Gottardi T, Braga RTV. Approaches for integration in system of systems. In: 4th International Workshop on Software Engineering for Systems-of-Systems. ACM; 2016. \pdoi{10.1145/2897829.2897835}
\item Silva E, Cavalcante E, Batista T. Refining missions to architectures in software-intensive systems-of-systems. In: IEEE/ACM Joint 5th International Workshop on Software Engineering for Systems-of-Systems and 11th Workshop on Distributed Software Development, Software Ecosystems and Systems-of-Systems (JSOS). IEEE; 2017. \pdoi{10.1109/jsos.2017.12}
\item Yun W, Shin D, Bae DH. Mutation analysis for system of systems policy testing. In: IEEE/ACM Joint 5th International Workshop on Software Engineering for Systems-of-Systems and 11th Workshop on Distributed Software Development, Software Ecosystems and Systems-of-Systems (JSOS). IEEE; 2017. \pdoi{10.1109/jsos.2017.9}
\item Motta RC, Oliveira KMD, Travassos GH. Rethinking interoperability in contemporary software systems. In: IEEE/ACM Joint 5th International Workshop on Software Engineering for Systems-of-Systems and 11th Workshop on Distributed Software Development, Software Ecosystems and Systems-of-Systems (JSOS). IEEE; 2017. \pdoi{10.1109/jsos.2017.5}
\item Alkhabbas F, Spalazzese R, Davidsson P. Emergent configurations in the Internet of Things as system of systems. In: IEEE/ACM Joint 5th International Workshop on Software Engineering for Systems-of-Systems and 11th Workshop on Distributed Software Development, Software Ecosystems and Systems-of-Systems (JSOS). IEEE; 2017. \pdoi{10.1109/jsos.2017.6}
\item Graciano Neto VV, Cavalcante E, El-Hachem J, Santos DS. On the interplay of Business Process Modeling and missions in systems-of-information systems. In: IEEE/ACM Joint 5th International Workshop on Software Engineering for Systems-of-Systems and 11th Workshop on Distributed Software Development, Software Ecosystems and Systems-of-Systems (JSOS). IEEE; 2017. \pdoi{10.1109/jsos.2017.7}
\item Mendes A, Loss S, Cavalcante E, Lopes F, Batista T. Mandala: An agent-based platform to support interoperability in systems-of-systems. In: 6th International Workshop on Software Engineering for Systems-of-Systems. ACM; 2018. \pdoi{10.1145/3194754.3194757}
\item Bouziat T, Camps V, Combettes S. A cooperative SoS architecting approach based on adaptive multi-agent systems. In: 6th International Workshop on Software Engineering for Systems-of-Systems. ACM; 2018. \pdoi{10.1145/3194754.3194756}
\item Baek YM, Song J, Shin YJ, Park S, Bae DH. A meta-model for representing system-of-systems ontologies. In: 6th International Workshop on Software Engineering for Systems-of-Systems. ACM; 2018. \pdoi{10.1145/3194754.3194755}
\item Neves VdO, Bertolino A, Angelis GD, Garcés L. Do we need new strategies for testing systems-of-systems? In: 6th International Workshop on Software Engineering for Systems-of-Systems. ACM; 2018. \pdoi{10.1145/3194754.3194758}
\item Bhardwaj N, Liggesmeyer P. A conceptual framework for safe reconfiguration in open system of systems. In: 6th International Workshop on Software Engineering for Systems-of-Systems. ACM; 2018. \pdoi{10.1145/3194754.3194759}
\item Olivero MA, Bertolino A, Dominguez-Mayo FJ, Escalona MJ, Matteucci I. Security assessment of systems of systems. In: IEEE/ACM 7th International Workshop on Software Engineering for Systems-of-Systems and 13th Workshop on Distributed Software Development, Software Ecosystems and Systems-of-Systems. IEEE; 2019. \pdoi{10.1109/sesos/wdes.2019.00017}
\item Song J, Torring JO, Hyun S, Jee E, Bae DH. Slicing executable system-of-systems models for efficient statistical verification. In: IEEE/ACM 7th International Workshop on Software Engineering for Systems-of-Systems and 13th Workshop on Distributed Software Development, Software Ecosystems and Systems-of-Systems. IEEE; 2019. \pdoi{10.1109/sesos/wdes.2019.00011}
\item Salvaneschi P. Emerging Structures in information systems: A SOS approach. In: IEEE/ACM 7th International Workshop on Software Engineering for Systems-of-Systems and 13th Workshop on Distributed Software Development, Software Ecosystems and Systems-of-Systems. IEEE; 2019. \pdoi{10.1109/sesos/wdes.2019.00012}
\item de Paula FC, Santos RPd. Systems Thinking as a resource for supporting accountability in system-of-information-systems: Exploring a Brazilian school case. In: IEEE/ACM 7th International Workshop on Software Engineering for Systems-of-Systems and 13th Workshop on Distributed Software Development, Software Ecosystems and Systems-of-Systems. IEEE; 2019. \pdoi{10.1109/sesos/wdes.2019.00014}
\item Carturan S, Goya D. Major challenges of systems-of-systems with cloud and DevOps - A financial experience report. In: IEEE/ACM 7th International Workshop on Software Engineering for Systems-of-Systems and 13th Workshop on Distributed Software Development, Software Ecosystems and Systems-of-Systems. IEEE; 2019. \pdoi{10.1109/sesos/wdes.2019.00010}
\item Teixeira PG, Lopes VHL, Santos RPd, Kassab M, Graciano Neto VV. The status quo of systems-of-information Systems. In: IEEE/ACM 7th International Workshop on Software Engineering for Systems-of-Systems and 13th Workshop on Distributed Software Development, Software Ecosystems and Systems-of-Systems. IEEE; 2019. \pdoi{10.1109/sesos/wdes.2019.00013}
\item Park S, Belay ZM, Bae DH. A simulation-based behavior analysis for MCI response system of systems. In: IEEE/ACM 7th International Workshop on Software Engineering for Systems-of-Systems and 13th Workshop on Distributed Software Development, Software Ecosystems and Systems-of-Systems. IEEE; 2019. \pdoi{10.1109/sesos/wdes.2019.00009}
\item Vargas IG, Nascimento DdL, Braga RTV. Fostering reuse by integration: A directed system of systems development case. In: 2020 IEEE International Conference on Software Architecture Companion. IEEE; 2020. \pdoi{10.1109/icsa-c50368.2020.00048}
\item Lopes SdS, Vargas IG, de Oliveira AL, Braga RTV. Risk management for system of systems: A systematic mapping study. In: 2020 IEEE International Conference on Software Architecture Companion. IEEE; 2020. \pdoi{10.1109/icsa-c50368.2020.00050}
\item Teixeira PG, Lebtag BGA, Santos RPd, et al. Constituent system design: A software architecture approach. In: 2020 IEEE International Conference on Software Architecture Companion. IEEE; 2020. \pdoi{10.1109/icsa-c50368.2020.00045}
\item Fonseca A, Sousa D, Chagas M, et al. Dealing with IoT defiant components. In: IEEE/ACM Joint 9th International Workshop on Software Engineering for Systems-of-Systems and 15th Workshop on Distributed Software Development, Software Ecosystems and Systems-of-Systems. IEEE; 2021. \pdoi{10.1109/sesos-wdes52566.2021.00009}
\item Ailane TM, Abboush M, Knieke C, Lawendy A, Rausch A. Toward formalizing the emergent behavior in Software Engineering. In: IEEE/ACM Joint 9th International Workshop on Software Engineering for Systems-of-Systems and 15th Workshop on Distributed Software Development, Software Ecosystems and Systems-of-Systems. IEEE; 2021. \pdoi{10.1109/sesos-wdes52566.2021.00010}
\item Pope M, Sillito J. Exploring non-functional coupling in systems of systems. In: 10th IEEE/ACM International Workshop on Software Engineering for Systems-of-Systems and Software Ecosystems. ACM; 2022. \pdoi{10.1145/3528229.3529383}
\item Borges MVdF, Rocha LS, Maia PHM. MicroGraphQL: A unified communication approach for systems of systems using microservices and GraphQL. In: 10th IEEE/ACM International Workshop on Software Engineering for Systems-of-Systems and Software Ecosystems. ACM; 2022. \pdoi{10.1145/3528229.3529381}
\item Michael J, Pfeiffer J, Rumpe B, Wortmann A. Integration challenges for digital twin systems-of-systems. In: 10th IEEE/ACM International Workshop on Software Engineering for Systems-of-Systems and Software Ecosystems. ACM; 2022. \pdoi{10.1145/3528229.3529384}
\item Heinrich J, Balduf F, Becker M, Adler R, Elberzhager F. Industry voices on Software Engineering challenges in dynamic systems of systems. In: IEEE/ACM 11th International Workshop on Software Engineering for Systems-of-Systems and Software Ecosystems. IEEE; 2023. \pdoi{10.1109/sesos59159.2023.00014}
\item Sjöberg P, Mendez D, Gorschek T. Contemporary challenges when developing cyber-Physical systems of systems - A case study. In: IEEE/ACM 11th International Workshop on Software Engineering for Systems-of-Systems and Software Ecosystems. IEEE; 2023. \pdoi{10.1109/sesos59159.2023.00012}
\end{enumerate}


\end{document}